\documentclass{article}

\PassOptionsToPackage{numbers, compress}{natbib}
\usepackage[preprint]{neurips_2025}

\usepackage[utf8]{inputenc} % allow utf-8 input
\usepackage[T1]{fontenc}    % use 8-bit T1 fonts
\usepackage{hyperref}       % hyperlinks
\usepackage{url}            % simple URL typesetting
\usepackage{booktabs}       % professional-quality tables
\usepackage{amsfonts}       % blackboard math symbols
\usepackage{nicefrac}       % compact symbols for 1/2, etc.
\usepackage{microtype}      % microtypography
\usepackage{xcolor}         % colors

\usepackage{microtype}
\usepackage{graphicx}
\usepackage{subfigure}
\usepackage{booktabs} % for professional tables
\usepackage{amsmath}
\usepackage{bm}
\usepackage{colortbl}
\usepackage{caption}
\usepackage{tablefootnote}
\usepackage{wrapfig}
\definecolor{verylightgray}{gray}{0.85}

\usepackage{multirow}
\usepackage{makecell}
\usepackage{adjustbox}
\newcommand{\para}[1]{\noindent\textbf{#1}}

\title{SpeechOp: Inference-Time Task Composition for Generative Speech Processing}

\author{%
  Justin Lovelace\thanks{Work done during an internship at Adobe. } \\
  Cornell University\\
  \texttt{jl3353@cornell.edu} \\
  \And
  Rithesh Kumar~~Jiaqi Su~~Ke Chen \\
  Adobe Research \\
  \texttt{\{ritheshk,jsu,kechen\}@adobe.com} \\
  \And
  Kilian Q Weinberger\\
  Cornell University\\
  \texttt{kqw4@cornell.edu} \\
  \And
  Zeyu Jin \\
  Adobe Research \\
  zejin@adobe.com
}

\begin{document}

\maketitle

\begin{abstract}
While generative Text-to-Speech (TTS) systems leverage vast ``in-the-wild" data to achieve remarkable success, speech-to-speech processing tasks like enhancement face data limitations, which lead data-hungry generative approaches to distort speech content and speaker identity. To bridge this gap, we present SpeechOp, a multi-task latent diffusion model that transforms pre-trained TTS models into a universal speech processor capable of performing a wide range of speech tasks and composing them in novel ways at inference time. By adapting a pre-trained TTS model, SpeechOp inherits a rich understanding of natural speech, accelerating training and improving S2S task quality, while simultaneously enhancing core TTS performance. Finally, we introduce Implicit Task Composition (ITC), a novel pipeline where ASR-derived transcripts (e.g., from Whisper) guide SpeechOp's enhancement via our principled inference-time task composition. ITC achieves state-of-the-art content preservation by robustly combining web-scale speech understanding with SpeechOp's generative capabilities. Audio samples are available at \url{https://justinlovelace.github.io/projects/speechop}.

\end{abstract}

\section{Introduction}
\label{sec:introduction}

% Main point: each person develops their own little model, not coming across crisply enough, 

Generative Text-to-Speech (TTS) systems now produce increasingly natural and expressive speech \citep{le2024voicebox, junaturalspeech}, largely due to their ability to leverage vast “in-the-wild” data (e.g., from audiobooks, podcasts \citep{GigaSpeech2021, pratap2020mls}). This scalability enables TTS models to learn robust speech representations across diverse acoustic conditions and speaker characteristics \citep{lee2024ditto, peng2024voicecraft}.

In contrast, speech-to-speech (S2S) processing tasks like enhancement, speaker separation, and foreground-background isolation face stricter data requirements, often needing paired degraded/clean speech, which is expensive to acquire at scale \citep{zen2019libritts}. Consequently, S2S models are typically trained on smaller, specialized datasets, often with simulated degradations \citep{su2021hifigan2}. This data scarcity can cause generative S2S approaches to distort original speaker identity and content—a critical issue where faithful preservation is paramount, e.g., in speech enhancement \citep{yang24h_interspeech, miipher}. These models often lack the rich speech understanding derived from vast, diverse datasets available to TTS.

\begin{figure*}[t]
    \centering
        \includegraphics[width=.9\textwidth]{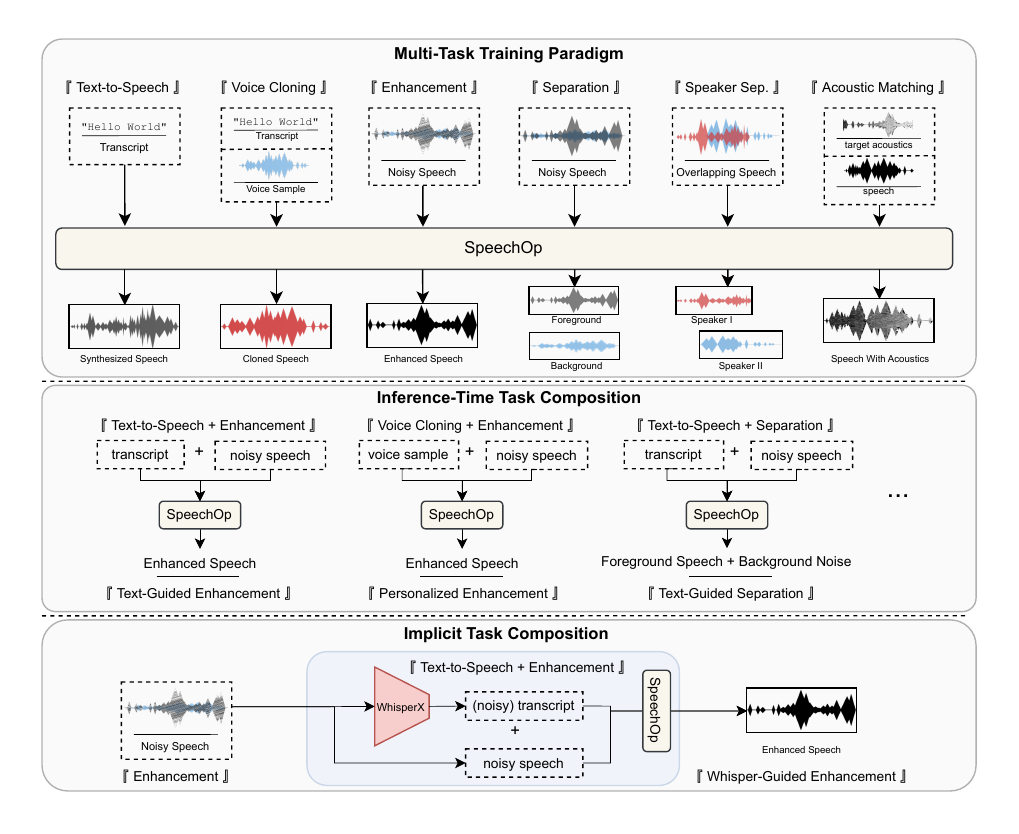}
        
    \caption{ Overview of SpeechOp's multi-task training (top), inference-time task composition capabilities (middle), and implicit task composition pipeline (bottom). The model is trained on six core speech tasks including text-to-speech, enhancement, and separation. At inference time, novel tasks can be composed by combining learned capabilities - for example, using transcripts to guide enhancement or personalizing enhancement with voice samples. In the implicit task composition pipeline, we use a state-of-the-art discriminative model (Whisper) to automatically transcribe noisy speech, then use the resulting transcript to guide SpeechOp's enhancement process.\label{fig:overview}}
    \vspace{-15pt}
\end{figure*}

To bridge this data gap, we present SpeechOp: a multi-task latent diffusion model that transforms pre-trained TTS models into a universal speech processor. SpeechOp performs a wide range of S2S tasks and allows their novel inference-time composition, leading to three key advancements (Figure~\ref{fig:overview}): (1) a flexible \textbf{multi-task} model enhancing core TTS quality, (2) inference-time \textbf{task composition} for unprecedented flexibility via our principled TC-CFG strategy, and (3) state-of-the-art S2S performance through \textbf{Implicit Task Composition (ITC)}, enabled by TC-CFG.

We make the following contributions: 

\textbf{1. A Flexible Multi-Task Model That Enhances TTS Capabilities:} SpeechOp, adapted from a pre-trained TTS model and fine-tuned on diverse S2S tasks (including TTS, enhancement, separation), not only becomes a versatile speech processor but also \textit{improves} its underlying TTS quality. By learning to handle varied acoustic manipulations, SpeechOp's TTS component generates more natural, higher-quality speech, validated by human listening studies.

\textbf{2. Inference-Time Task Composition (TC-CFG):} 
For instance, if speech content is obscured, SpeechOp can combine its enhancement capabilities with TTS content guidance to both enhance acoustics and \textit{re-synthesize} the obscured portion. Crucially, our novel TC-CFG guidance strategy (Section~\ref{sec:task_composition}) enables this powerful composition at \textit{inference-time} without requiring joint training.  

\textbf{3. State-of-the-Art Speech Processing through Implicit Task Composition (ITC):} SpeechOp achieves state-of-the-art content preservation via ITC. Traditional transcript-conditioned S2S models suffer from scarce paired noisy-clean-transcript data and the propagation of ASR errors. ITC overcomes these by robustly integrating ASR-derived transcripts (e.g., from Whisper \citep{radford2023robust, bain2022whisperx}) using our TC-CFG inference-time composition. This principled approach, with its tunable ``guidance strength,'' allows balancing content restoration (more like TTS) and acoustic fidelity (more like enhancement) based on the situation, achieving superior content fidelity over specialized enhancement methods.

\section{Background: Diffusion Models}
\label{sec:background}

We introduce latent diffusion models following recent formulations \citep{ho2020denoising, kingma2023understanding, rombach2021highresolution}. Given data drawn from an unknown distribution $q(\mathbf{x})$, our goal is to learn a generative model $p_\theta(\mathbf{x})$ that approximates this distribution. 

\para{Forward process.}
The forward process defines a gradual transition from the latent distribution to a Gaussian distribution through a sequence of increasingly noisy latent variables $\mathbf{z}_t$ for timesteps $t\in[0,1]$. This Gaussian diffusion process defines the conditional distribution $q(\mathbf{z}_{0,\ldots,1}|\mathbf{x})$. For every $t\in[0,1]$, the marginal $q(\mathbf{z}_{t}|\mathbf{x})$ is given by:
\[
\mathbf{z}_t = \alpha_{t} \mathbf{x} + \sigma_{t} \bm{\epsilon}, \quad \text{where} \quad \bm{\epsilon} \sim \mathcal{N}(\mathbf{0}, \mathbf{I})
\]
We use the variance-preserving formulation where $\sigma_{t}^2=1-\alpha_{t}^2$. The noise schedule $\alpha_t\in [0,1]$ is a strictly monotonically decreasing function that starts with the original latent ($\mathbf{z}_0 \approx \mathbf{x}$) and ends with approximately Gaussian noise ($q(\mathbf{z}_1) \approx \mathcal{N}(\mathbf{z}_1; \mathbf{0}, \mathbf{I})$).

\para{Generative model.}
Given the score function $\nabla_{\mathbf{z}
}\log q_t(\mathbf{z}
)$, or the gradient of the log probability density function, we can reverse the forward process exactly. 
Diffusion models utilize a neural network learn to estimate the score function, $\mathbf{s}_\theta(\mathbf{z};\lambda)\approx\nabla_{\mathbf{z}}\log q_t(\mathbf{z})$, and use the estimated score function to approximately reverse the forward process. If $\mathbf{s}_\theta(\mathbf{z};\lambda)\approx\nabla_{\mathbf{z}}\log q_t(\mathbf{z})$, then our generative distribution is close to the true distribution. This enables us to draw samples from a Gaussian distribution $\mathbf{z}_1 \sim p(\mathbf{z}_1)$, and approximately solve the reverse diffusion process using the estimated score $\mathbf{s}_\theta(\mathbf{z};\lambda)$.

\para{Training objective.} 
We train the score network using a denoising score matching (DSM) loss \citet{song2019generative} over all data points $\mathbf{x}\sim\mathcal{D}$ and noise levels:
\begin{align*}
    \mathcal{L}_{\text{DSM}}(\mathbf{x}) = \mathbb{E}_{t,\mathbf{x}, \bm{\epsilon}}[{w}(\lambda_t)\cdot \lVert\mathbf{s}_\theta(\mathbf{z}_t;\lambda) - \nabla_{\mathbf{z}_t}\log q(\mathbf{z}_t|\mathbf{x})\rVert_2^2],
\end{align*}
where $w(\lambda_t)$ weights different noise levels during training. Following best practices \citep{salimans2022progressive}, we adopt the velocity parameterization, $\mathbf{v} = {\alpha_t}\bm{\epsilon} - \sigma_t \mathbf{x}$, for our network output to ensure training stability.

\section{Related Work}
\label{sec:related_work}

Text-to-speech (TTS) systems \citep{le2024voicebox, shennaturalspeech} excel due to vast "in-the-wild" data, unlike data-limited speech-to-speech (S2S) tasks like enhancement \citep{miipher} and separation, which often require scarce paired recordings. While multi-task autoregressive models have been developed \citep{wang2024speechx}, they often lack inference-time compositionality. Diffusion models, especially latent ones, offer high-quality synthesis \citep{shennaturalspeech, le2024voicebox} and inherent composability \citep{liu2022compositional}, which SpeechOp leverages.

SpeechOp adapts pre-trained TTS models for S2S tasks and novel inference-time composition. Recent work like Fugatto \citep{valle2025fugatto} also explores multi-task audio generation. However, SpeechOp's principled task composition (TC-CFG, Section~\ref{sec:task_composition}) provides superior control and performance in S2S tasks like enhancement compared to Fugatto's score averaging (Section~\ref{sec:results}), enabling effective combination of operations like enhancement and TTS.
While foundational models like UniAudio \citep{yang2023uniaudio} pursue broad task coverage from scratch, and SpeechFlow \citep{liu2024generative} investigates new pre-training schemes, SpeechOp focuses on efficiently adapting \textbf{existing} TTS models. Crucially, we introduce Implicit Task Composition (ITC), which uniquely integrates ASR models (e.g., Whisper) via TC-CFG for robust content preservation. Our primary aim is not maximizing task variety, but demonstrating how TTS pre-training and sophisticated composition can address S2S data scarcity and improve performance on established operations.

% \begin{wrapfigure}{r}{0.5\textwidth}
% \centering
% \centering
% \includegraphics[width=\linewidth]{figures/validation_mse_comparison_unified.pdf}
% \label{fig:validation_s2s}
% \vspace{-15pt}
% \caption{Validation loss curves demonstrating accelerated convergence when initializing the diffusion transformer from a TTS backbone. TTS initialization reduces training time by 4$\times$ for speech enhancement and 8$\times$ for speaker separation.}

% % \vspace{1em}

% \captionof{table}{Impact of Initialization on Speech Processing.}
% \label{tab:init_performance}
% \resizebox{.5\textwidth}{!}{  
% \begin{tabular}{lcccccc}
% \toprule
% & \multicolumn{5}{c}{\textbf{Speaker Separation}} \\
% \cmidrule(lr){2-6}
% Init & SI-SDRi $\uparrow$ & MCD $\downarrow$ & SpBS $\uparrow$ & WER $\downarrow$ & Val MSE $\downarrow$ \\
% \midrule
% Rand & -3.99 & 22.95 & .825 & 17.8 & .358 \\
% TTS & -1.38 & 4.46 & .906 & 8.5 & .336 \\
% \midrule
% & \multicolumn{5}{c}{\textbf{Speech Enhancement}} \\
% \cmidrule(lr){2-6}
% Init & PESQ $\uparrow$ & MCD $\downarrow$ & SpBS $\uparrow$ & WER $\downarrow$ & Val MSE $\downarrow$ \\
% \midrule
% Rand & 2.07 & 4.76 & .900 & 8.1 & 0.289 \\
% TTS & 2.10 & 4.69 & .910 & 8.1 & 0.276 \\
% \bottomrule
% \end{tabular}
% }
% \end{wrapfigure}

\begin{figure}[ht] % Using [ht] for more placement flexibility, [t] is also fine.
    \centering
    \begin{minipage}[t]{0.47\linewidth} % Use \linewidth for width relative to minipage
        \centering
        % Wrap image in adjustbox to ensure its true top is used for alignment
        \adjustbox{valign=t, margin=0pt}{%
            \includegraphics[width=\linewidth]{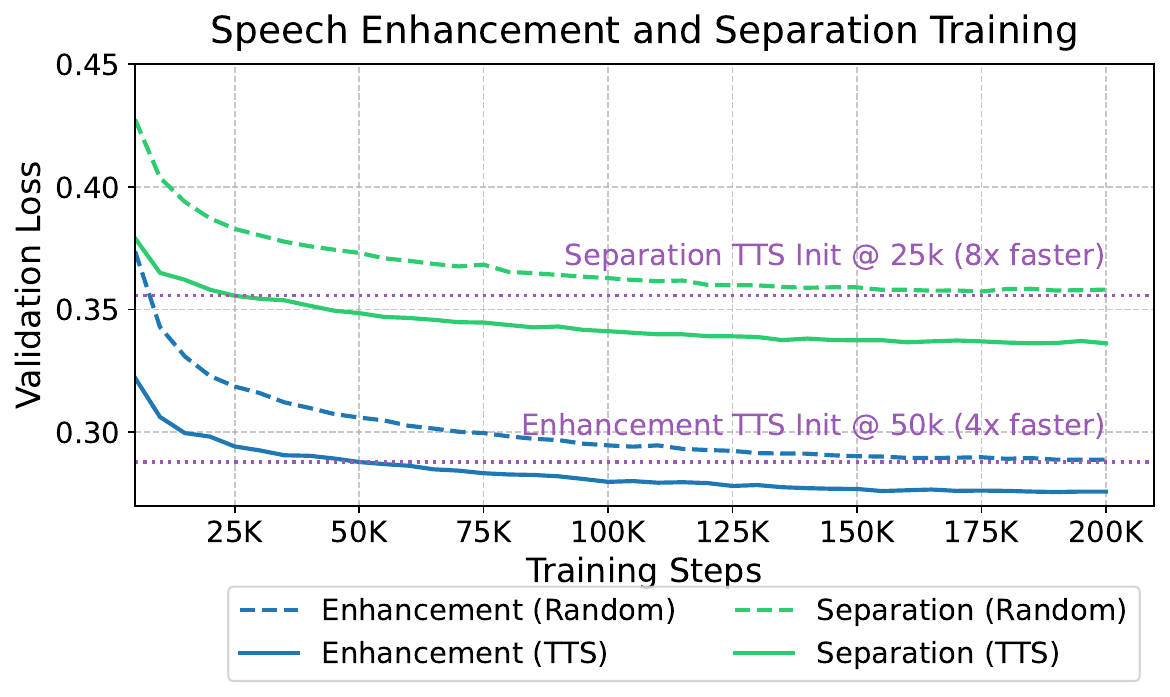}%
        }
        % --- REMOVE CAPTION AND LABEL FROM HERE ---
    \end{minipage}
    \hfill % Adds flexible space between the minipages
    \begin{minipage}[t]{0.51\linewidth} % Use \linewidth for width relative to minipage
        \centering
        \adjustbox{valign=t, margin=0pt}{%
            \resizebox{\linewidth}{!}{%
            \begin{tabular}{lcccccc}
            \toprule
            & \multicolumn{5}{c}{\textbf{Speaker Separation}} \\
            \cmidrule(lr){2-6}
            Init & SI-SDRi $\uparrow$ & MCD $\downarrow$ & SpBS $\uparrow$ & WER $\downarrow$ & Val MSE $\downarrow$ \\
            \midrule
            Rand & -3.99 & 22.95 & .825 & 17.8 & .358 \\
            TTS & -1.38 & 4.46 & .906 & 8.5 & .336 \\
            \bottomrule
            \vspace{.5em} \\ % This adds vertical space within the table
            \toprule
            & \multicolumn{5}{c}{\textbf{Speech Enhancement}} \\
            \cmidrule(lr){2-6}
            Init & PESQ $\uparrow$ & MCD $\downarrow$ & SpBS $\uparrow$ & WER $\downarrow$ & Val MSE $\downarrow$ \\
            \midrule
            Rand & 2.07 & 4.76 & .900 & 8.1 & 0.289 \\
            TTS & 2.10 & 4.69 & .910 & 8.1 & 0.276 \\
            \bottomrule
            \end{tabular}%
            }%
        }
        % --- REMOVE CAPTIONOF AND LABEL FROM HERE ---
    \end{minipage}
    % --- ADD UNIFIED CAPTION AND LABEL HERE ---
    \caption{Impact of TTS initialization on speech processing tasks.
    (\textbf{Left}) Validation loss curves demonstrating accelerated convergence with TTS initialization. Training time is reduced by 4$\times$ for enhancement and 8$\times$ for separation.
    (\textbf{Right}) Performance metrics for speaker separation and speech enhancement, comparing random initialization (Rand) with TTS initialization (TTS).
    }
    \label{fig:s2s_init_comparison_unified} % A single label for the entire figure
    \vspace{-15pt}
\end{figure} 
\section{TTS Pre-training Improves Speech Processing Tasks}
\label{sec:transfer}
To motivate our multi-task framework, we first examine the benefits of initializing single-task speech enhancement and speaker separation models from a pre-trained DiT TTS backbone \cite{peebles2022scalable, lee2024ditto}. 
Figure~\ref{fig:s2s_init_comparison_unified} (Left) shows that TTS initialization dramatically accelerates convergence, achieving comparable validation loss with 4$\times$ fewer steps for enhancement and 8$\times$ fewer for separation versus random initialization. The significant speedup for separation, a complex multi-speaker task, particularly highlights the value of the speech understanding capabilities inherited from TTS pre-training.

Beyond faster training, TTS pre-training yields substantial performance gains (Figure~\ref{fig:s2s_init_comparison_unified} Right). Speaker separation benefits most, with TTS initialization leading to markedly improved SI-SDRi, MCD, and WER (8.5\% vs 17.8\%), and eliminating artifacts present in randomly initialized models that struggle with content disentanglement. Speech enhancement also sees improvements in PESQ, MCD, and SpeechBERTScore. 
These results demonstrate the broad advantages of TTS pre-training—accelerated convergence and enhanced performance across diverse S2S tasks, especially those requiring deep speech understanding. This motivates SpeechOp, our multi-task framework leveraging TTS pre-training for high-quality, versatile speech processing.

\section{SpeechOp}
\label{sec:methodology} 

We present the tasks explored in this work in Figure~\ref{fig:overview}. These tasks provide complementary capabilities that are composable via our diffusion framework for applications like transcript-guided isolation. For speaker separation, which requires a speaker prompt to identify the target speaker, we provide a disjoint speech sample to disambiguate the target speaker. For foreground/background separation, we parameterize them as two separate tasks.

\para{SpeechOp Architecture.}
\label{sec:speechop_architecture}
SpeechOp is built on a latent diffusion framework \citep{rombach2021highresolution} that operates with compressed audio representations. Rather than working with raw waveforms, we first compress the audio using a DAC variational autoencoder \cite{kumar2024high} (details in Appendix), allowing our model to efficiently process and generate speech in a lower-dimensional latent space.

\begin{wrapfigure}{r}{0.5\linewidth}
    \centering
    \includegraphics[width=\linewidth]{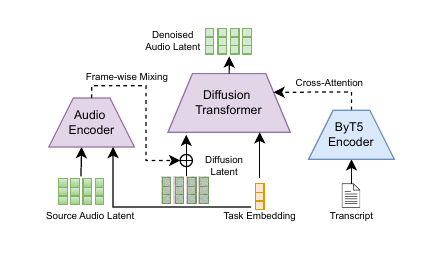}
    \caption{SpeechOp Architecture Overview. \label{fig:arch}}
\end{wrapfigure}

As shown in Figure~\ref{fig:arch}, SpeechOp's core architecture consists of a Diffusion Transformer (DiT) \citep{peebles2023scalable} that serves as the denoising network, adapted to handle both text-to-speech and speech-to-speech tasks. The model processes text transcripts for TTS and source audio (like noisy speech) for speech-to-speech tasks, with a learnable Task Embedding that conditions model behavior. Training proceeds in two stages: TTS pre-training followed by multi-task training to enable speech-to-speech capabilities.

\para{Text-to-Speech Pathway.}
For TTS, SpeechOp (Figure~\ref{fig:arch}, right) processes a text transcript. We extract transcript representations with a frozen, pre-trained ByT5-base encoder \cite{xue2022byt5, lovelace2024simpletts}. ByT5's character-level representations capture phonetic information crucial for natural speech. The DiT is conditioned on the ByT5 embeddings via cross-attention, dynamically aligning text and audio frames and guiding denoising based on text content. For our Diffusion Transformer (DiT) architecture, we incorporate design choices from recent TTS systems \citep{lee2024ditto, lovelace24_interspeech} (full details in the Appendix).

To enable speaker-prompted generation and speech editing, we train our model to perform inpainting for 75\% of samples \cite{le2024voicebox}. After adding noise from the forward diffusion process, we replace a random segment of the latent with the clean, target segment. We additionally sum a learnable binary embedding at the input layer to distinguish clean from noisy frames. The network will then learn to extrapolate speaker and speech properties from the ground-truth region to denoise the noisy speech. For half of our inpainting samples, we replace the initial segment (simulating voice prompts). In the other half, we noise only the middle section, replacing the start and end of the utterance with clean speech to simulate speech editing. For sampling the relative duration, we follow \citet{lovelace24_interspeech} and use a Beta distribution with a mode of $.01$ and a concentration of $5$ to emphasize challenging cases with short prompts.

\para{Speech-to-Speech Pathway.}
To handle speech-to-speech tasks like enhancement and separation (Figure~\ref{fig:arch}, left), SpeechOp introduces a dedicated Audio Encoder to process source audio such as noisy speech. This encoder adopts the same DiT architecture as the main model but starts with random initialization. Since speech-to-speech tasks inherently maintain frame-level alignment between source and target audio, we implement a straightforward frame-wise mixing approach rather than us a complex alignment mechanism. Specifically, the Audio Encoder's output representations are directly added to the Diffusion Latent before processing by the Diffusion Transformer, allowing direct incorporation of source audio information during denoising.
To handle different speech-to-speech tasks, we use a learnable Task Embedding that conditions both the Audio Encoder and Diffusion Transformer. This shared embedding provides task-specific guidance to both components based on the desired operation (enhancement, separation, etc.).

Some speech-to-speech tasks require additional input prompts. For example, speaker separation needs a reference speech sample to identify the target speaker, while acoustic matching needs an example of the target acoustics. In these cases, we prepend the prompt to both the source audio and noisy latent to maintain frame-wise alignment. For tasks that typically don't use prompts (like enhancement), we unmask the latent's initial segment in 10\% of training instances to enable transfer learning with speaker-prompted TTS. These prompt durations follow the same Beta distribution used for TTS inpainting.

\para{Multi-Task Fine-Tuning.}
\label{sec:multi_task_finetuning}
SpeechOp uses a two-stage training approach. After initial TTS pre-training, we conduct multi-task fine-tuning where both the Audio Encoder and pre-trained DiT backbone are jointly optimized. During this stage, we sample TTS and speech-to-speech (S2S) data with equal frequency. Within the S2S samples, we apply selective upsampling - tripling the frequency of enhancement and speaker separation examples since these are the most challenging tasks. This two-stage strategy efficiently adapts the TTS model into our multi-task SpeechOp model.

\para{Diffusion Training.}
\label{subsec:diffusion_process}
During training, we sample noise levels using a shifted cosine schedule (s=0.5) \citep{hoogeboom2023simple}, following \citet{lovelace2024simpletts}. We employ the Sigmoid diffusion loss weighting from \citet{hoogeboom2024simpler} with a bias of -2.5 to concentrate training on perceptually relevant noise levels. To enable classifier-free guidance during inference, we randomly drop conditioning information (source audio and transcript) 10\% of the time during training \citep{ho2022classifier}.

\section{Inference-Time Task Composition}
\label{sec:task_composition}
The ability to compose speech operations—such as simultaneously enhancing noisy speech while restoring its content via text—represents a powerful capability for speech processing. 
Text-guided generation can help produce a plausible, high-fidelity version of content that is otherwise not recoverable from complex acoustic situations, such as intense noise and reverberation encountered in speech enhancement. Similarly, in speaker separation, the text of spoken content could provide important contextual cues for disentangling speakers.
Nonetheless, achieving an effective composition of tasks poses significant technical challenges.

Prior work, including Fugatto in the audio domain, typically computes a weighted average of score functions to compose operations\cite{liu2022compositional, valle2025fugatto}, like for enhancement and TTS:
\begin{equation}
\mathbf{s}^{\text{avg}}_{\theta}(\mathbf{z}_t | y, w) = (1-\alpha) \mathbf{s}^{\text{enh}}_{\theta}(\mathbf{z}_t|y) + \alpha \mathbf{s}^{\text{tts-prior}}_{\theta}(\mathbf{z}_t|w)
\label{eq:baseline_averaging}
\end{equation}
Here, $\mathbf{s}^{\text{enh}}_{\theta}(\mathbf{z}_t|y)$ is the score from an enhancement model conditioned on noisy audio $y$, and $\mathbf{s}^{\text{tts}}_{\theta}(\mathbf{z}_t|w)$ represents a score function derived from a TTS model aiming to generate speech for transcript $w$. While straightforward, this approach poses a fundamental limitation: it combines the generative priors of different tasks. For speech enhancement with TTS guidance, direct averaging allows the TTS model's broad acoustic prior (learned from diverse data for generation) to corrupt the enhancement model's focused studio-quality prior (learned for reconstruction), degrading output quality.

To address this challenge, we propose decomposing the desired score function $\nabla_{\mathbf{z}_t} \log p(\mathbf{z}_t|y,w)$ into task-specific components. Using Bayes' rule and a conditional independence assumption (transcript $w$ is independent of noisy audio $y$ given latent $\mathbf{z}_t$; detailed derivation in the Appendix), we arrive at:
\begin{equation}
    \nabla_{\mathbf{z}_t} \log p(\mathbf{z}_t|y,w) = \nabla_{\mathbf{z}_t} \log p(\mathbf{z}_t|y) + \nabla_{\mathbf{z}_t} \log p(w|\mathbf{z}_t).
    \label{eq:intermediate_score_function_main}
\end{equation}
This decomposition yields two complementary terms: an enhancement score $\nabla_{\mathbf{z}_t} \log p(\mathbf{z}_t|y)$ that guides acoustic quality based on the input $y$, and a discriminative guide $\nabla_{\mathbf{z}_t} \log p(w|\mathbf{z}_t)$. This second term leverages a TTS model not for its generative prior, but for its ability to \textit{discriminate} whether a latent $\mathbf{z}_t$ is likely to produce content matching transcript $w$. This term guides the latent towards speech aligned with the transcript without imposing the TTS model's full acoustic prior.

\para{Implementation via Classifier-Free Guidance.}
We implement this decomposition using classifier-free guidance (CFG) \cite{ho2022classifier} to approximate the discriminative signal $\nabla_{\mathbf{z}_t} \log p(w|\mathbf{z}_t)$:
\begin{align}
    \nabla_{\mathbf{z}_t} \log p(w|\mathbf{z}_t) &\approx \gamma \left( \mathbf{s}^{\text{tts}}_{\theta}(\mathbf{z}_t|w) - \mathbf{s}^{\text{tts}}_{\theta}(\mathbf{z}_t) \right)
\end{align}
where $\mathbf{s}^{\text{tts}}_{\theta}(\mathbf{z}_t|w)$ is the score of a TTS model conditioned on transcript $w$, $\mathbf{s}^{\text{tts}}_{\theta}(\mathbf{z}_t)$ is its unconditional score, and $\gamma$ is a guidance scale. Substituting this into Eq.~\eqref{eq:intermediate_score_function_main}, our final composed score is:
\begin{equation}
    \mathbf{s}^{\text{CFG}}_{\theta}(\mathbf{z}_t|y,w) \approx \mathbf{s}^{\text{enh}}_{\theta}(\mathbf{z}_t|y) + \gamma \left( \mathbf{s}^{\text{tts}}_{\theta}(\mathbf{z}_t|w) - \mathbf{s}^{\text{tts}}_{\theta}(\mathbf{z}_t) \right).
    \label{eq:final_score_function_main}
\end{equation}
This formulation, which we term Task-Composition Classifier-Free Guidance (TC-CFG), preserves the strengths of both tasks. The enhancement term maintains acoustic quality and speaker characteristics. The CFG-derived discriminative term provides content alignment by isolating text-specific guidance, avoiding the pitfalls of directly mixing generative TTS priors with the enhancement prior.

\begin{figure*}[t!]
    \centering
    \includegraphics[width=\textwidth]{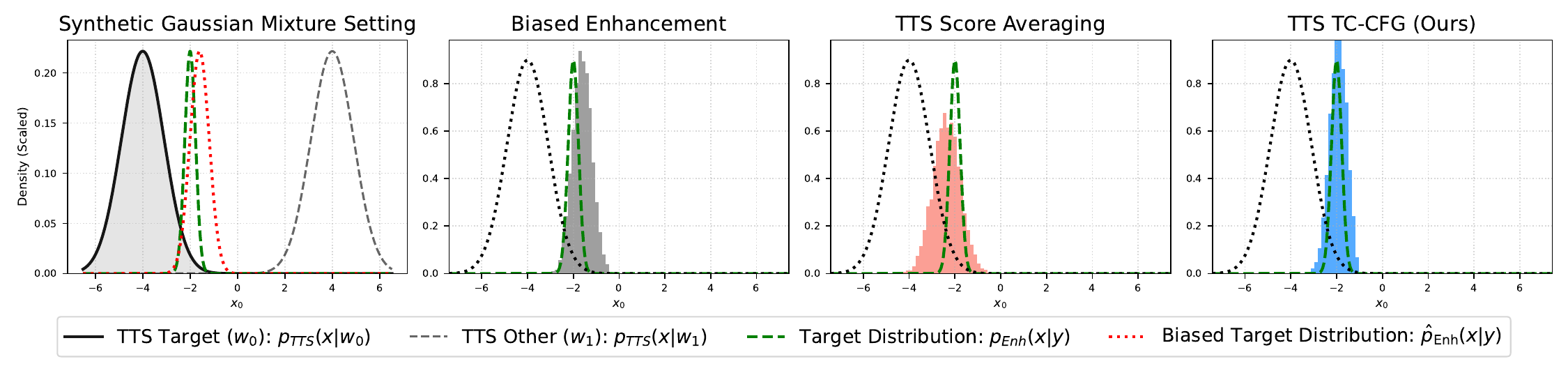}
    \caption{\textbf{A 1D toy simulation illustrating different task composition methods.} 
    \textbf{(a)} The setup, showing a bimodal TTS prior (target $w_0$ and other $w_1$), an ideal sharp target distribution ($p_{\text{Enh}}$), and a biased enhancement model ($\hat{p}_{\text{Enh}}$) whose output is misaligned. 
    \textbf{(b)} Samples from the unguided biased model.
    \textbf{(c)} Samples using score averaging (Eq.~\eqref{eq:baseline_averaging}).
    \textbf{(d)} Samples using our TC-CFG method (Eq.~\eqref{eq:final_score_function_main}).
    \label{fig:toy_sim_combined}}
    \vspace{-10pt}
\end{figure*}

\para{Synthetic Simulation.} To illustrate the downsides of score averaging and motivate our TC-CFG approach, we use a 1D Gaussian mixture simulation (Figure~\ref{fig:toy_sim_combined}, full details in appendix).
The setup (Figure~\ref{fig:toy_sim_combined}a) features a bimodal TTS prior, a sharp ideal enhanced distribution $p_{\text{Enh}}$ for the target word $w_0$, and an imperfect (biased) enhancement model $\hat{p}_{\text{Enh}}$ whose output for $w_0$ is misaligned.
Without guidance, the biased model's samples are incorrect (Figure~\ref{fig:toy_sim_combined}b). However, combining the biased enhancement score function with the TTS score function can potentially correct for content errors.

Score averaging (Figure~\ref{fig:toy_sim_combined}c) pulls samples towards the $w_0$ TTS mode. However, because this mixes in the broad TTS prior, the result is a "smeared" distribution that compromises the target's distribution.
In contrast, our TC-CFG approach (Figure~\ref{fig:toy_sim_combined}d) incorporates discriminative TTS guidance (via $\nabla_{x_t} \log p_{\text{TTS}}(w_0|x_t)$) to steer sampling. This shifts the sampled distribution to satisfy the discriminative signal without compromising the enhancement prior. We demonstrate the benefits of this approach extend to the real-world setting of composing enhancement with TTS synthesis in \autoref{sec:results}.

\section{Experimental Setup}
\label{sec:experimental_setup}
SpeechOp integrates a 20-layer Diffusion Transformer (DiT, 419M parameters) with an 8-layer audio encoder (71M parameters). We compare it against strong baselines for speech enhancement and speaker separation.

\para{Training Data.} For TTS, we combine MLS English (~44k hours) \citep{pratap2020mls} for longer utterances (10-20s) and Libri-TTS (585 hours) \citep{zen2019libritts} for shorter segments (<10s), improving robustness. All audio is resampled to 48kHz and transcripts lowercased. For S2S tasks, we use LibriTTS-R \citep{koizumi23_interspeech} for clean speech and simulate degradations using established noise/impulse response datasets and pipelines \citep{yang24h_interspeech}, creating 5s paired instances. Further dataset details are in the Appendix.

\para{Tasks and Baselines.}
\textbf{Text-to-Speech (TTS):} Evaluated on LibriSpeech test-clean \citep{librispeech} against contemporary end-to-end TTS systems \citep{le2024voicebox, chen2024images, lee2024ditto}. Speech editing is evaluated on the LibriTTS portion of RealEdit \citep{peng2024voicecraft}.
\textbf{Speech Enhancement (SE):} Baselines include waveform (StoRm \citep{lemercier2023storm}) and diffusion-based (SGMSE+ \citep{richter2023speech}, Miipher+WavLM \citep{koizumi2023miipher, yang24h_interspeech}) models, and GAN-based HiFi-GAN-2 \citep{su2021hifigan2}.
\textbf{Speaker Separation (SS):} We compare against Sepformer variants \citep{subakan2021attention, chen24h_interspeech}, including those trained on WHAMR! \citep{maciejewski2020whamr} and with acoustic-content simulation.

\para{Evaluation Metrics.}
\label{subsec:evaluation_metrics}
We assess SpeechOp on four dimensions:
\textbf{Subjective Quality:} Mean Opinion Scores (MOS, 1-5 scale) from listening tests on Prolific \citep{prolific}.
\textbf{Signal Similarity:} PESQ (perceived quality), MCD (spectral distance, lower is better), and SI-SDRi (separation distortion improvement) \cite{rouxwe2019SISDR}.
\textbf{Neural Similarity:} WavLM-TDCNN \citep{Chen2021WavLM} for speaker similarity (SIM) and SpeechBERTScore (SpBS) \cite{saeki2024speechbertscore} for semantic alignment.
\textbf{Content Accuracy:} Word Error Rate (WER) via HuBERT-L \cite{hsu2021hubert} (TTS) and WhisperX (large-v2) \cite{radford2023robust, bain2022whisperx} (other tasks).

\begin{table*}[t]
\caption{Zero-Shot Text-to-Speech Evaluation. MOS metrics evaluate different aspects: MOS-Q (Quality), MOS-N (Naturalness), MOS-VS (Voice Similarity), and MOS-SS (Style Similarity). Models in a different parameter regime are displayed in \textcolor{gray}{gray}.}
\label{tab:zeroshot_tts_results}
\centering
\resizebox{\linewidth}{!}{
\begin{tabular}{lcccccccc}
\toprule
Model & Params & Training Data & WER $\downarrow$ & SIM $\uparrow$ & MOS-Q $\uparrow$ & MOS-N $\uparrow$ & MOS-VS $\uparrow$ & MOS-SS $\uparrow$ \\
\midrule
Ground Truth & — & — & 2.19 & 0.67 & 4.24 {\tiny$\pm$ 0.06} & 4.16 {\tiny$\pm$ 0.06} & 3.79 {\tiny$\pm$ 0.06} & 3.60 {\tiny$\pm$ 0.06} \\
\midrule
\textcolor{gray}{DiTTo-TTS} \citep{lee2024ditto} & \textcolor{gray}{740M} & \textcolor{gray}{$\sim$56k hrs} & \textcolor{gray}{2.56} & \textcolor{gray}{.62} & \textcolor{gray}{4.16 {\tiny$\pm$ 0.04}} & \textcolor{gray}{4.14 {\tiny$\pm$ 0.04}} & \textcolor{gray}{4.17 {\tiny$\pm$ 0.04}} & \textcolor{gray}{4.02 {\tiny$\pm$ 0.04}} \\
\textcolor{gray}{VoiceCraft} \citep{peng2024voicecraft} & \textcolor{gray}{830M} & \textcolor{gray}{$\sim$69k hrs} & \textcolor{gray}{6.32} & \textcolor{gray}{.61} & \textcolor{gray}{3.66 {\tiny$\pm$ 0.04}} & \textcolor{gray}{3.65 {\tiny$\pm$ 0.05}} & \textcolor{gray}{3.43 {\tiny$\pm$ 0.05}} & \textcolor{gray}{3.38 {\tiny$\pm$ 0.05}} \\
CLaM-TTS \citep{kimclam} & 584M & $\sim$56k hrs & 5.11 & .49 & 3.67 {\tiny$\pm$ 0.04} & 3.70 {\tiny$\pm$ 0.04} & 3.69 {\tiny$\pm$ 0.05} & 3.54 {\tiny$\pm$ 0.05} \\
XTTS \citep{casanova2024xtts} & 482M & $\sim$17k hrs & 4.93 & .49 & 3.76 {\tiny$\pm$ 0.04} & 3.66 {\tiny$\pm$ 0.05} & 3.28 {\tiny$\pm$ 0.05} & 3.27 {\tiny$\pm$ 0.05} \\
\midrule
TTS Baseline (Ours) & 419M & $\sim$45k hrs & 3.32 & .48 & 3.65 {\tiny$\pm$ 0.05} & 3.56 {\tiny$\pm$ 0.05} & 3.31 {\tiny$\pm$ 0.05} & 3.25 {\tiny$\pm$ 0.05} \\
SpeechOp (Ours) & 419M & $\sim$45k hrs& 3.57 & .53 & 3.86 {\tiny$\pm$ 0.04} & 3.69 {\tiny$\pm$ 0.05} & 3.67 {\tiny$\pm$ 0.05} & 3.58 {\tiny$\pm$ 0.05} \\
$\Delta$ from Multi-Task Training &  — & — & \textcolor{red}{+0.25} & \textcolor{blue}{+.05} & \textbf{\textcolor{blue}{+0.22{\tiny$\pm$ 0.06}}} & \textbf{\textcolor{blue}{+0.13{\tiny$\pm$ 0.07}}} & \textbf{\textcolor{blue}{+0.36{\tiny$\pm$ 0.07}}} & \textbf{\textcolor{blue}{+0.32{\tiny$\pm$ 0.07}}} \\
\bottomrule
\end{tabular}}
\caption{Speech Editing Evaluation. }
\label{tab:speech_editing_results}
\centering
\resizebox{\linewidth}{!}{
\begin{tabular}{lccccccc}
\toprule
Model & Params & Training Data & WER $\downarrow$  & MOS-Q $\uparrow$ & MOS-N $\uparrow$ & MOS-VS $\uparrow$ & MOS-SS $\uparrow$ \\
\midrule
Ground Truth & — & — & 16.2 & 4.33 {\tiny$\pm$ 0.04} & 4.40 {\tiny$\pm$ 0.03} & 4.66 {\tiny$\pm$ 0.03} & 4.63 {\tiny$\pm$ 0.03} \\
\midrule
\textcolor{gray}{VoiceCraft} \citep{peng2024voicecraft} & \textcolor{gray}{830M} & \textcolor{gray}{$\sim$69k hrs} & \textcolor{gray}{16.3}& \textcolor{gray}{3.62 {\tiny$\pm$ 0.04}} & \textcolor{gray}{3.99 {\tiny$\pm$ 0.04}} & \textcolor{gray}{4.12 {\tiny$\pm$ 0.04}} & \textcolor{gray}{4.01 {\tiny$\pm$ 0.04}} \\
\midrule
TTS Baseline (Ours) & 419M & $\sim$45k hrs & 16.4 & 4.18 {\tiny$\pm$ 0.04} & 4.23 {\tiny$\pm$ 0.04} & 4.45 {\tiny$\pm$ 0.03} & 4.23 {\tiny$\pm$ 0.04} \\
SpeechOp (Ours) & 419M & $\sim$45k hrs& 15.9 & 4.15 {\tiny$\pm$ 0.04} & 4.19 {\tiny$\pm$ 0.04} & 4.48 {\tiny$\pm$ 0.03} & 4.25 {\tiny$\pm$ 0.03} \\
\bottomrule
\end{tabular}}
\vspace{-15pt}
\end{table*}
\begin{table}[ht]
    \centering
    \caption{Speech Enhancement Results. (\textbf{Left}) Quantitative metrics. (\textbf{Right}) Subjective MOS scores with standard error.}
    \label{tab:enhancement_results_combined} % Single label for the combined tables
    \begin{minipage}[t]{0.65\linewidth}
        \centering
        % \captionof{table}{Speech Enhancement Results (Quant.)} % REMOVED
        % \label{tab:enhancement_results_quant} % REMOVED
        \resizebox{\linewidth}{!}{
        \begin{tabular}{lcccc}
        % ... quantitative table content ...
        \toprule
        Model & PESQ $\uparrow$ & MCD $\downarrow$ & SpBS $\uparrow$ & WER $\downarrow$  \\
        \midrule
        Noisy Source Audio & 1.12 & 11.22 & .888 & 3.3  \\
        \midrule
        Storm & 1.61 & 6.36 & .883 & 7.0  \\
        Miipher & 1.44 & 5.15 & .898 & 7.0  \\
        SGMSE+ & 1.98 & 5.28 & .923 & 5.7  \\
        HiFi-GAN-2 & 2.23 & 4.40 & .934 & 5.4  \\
        \midrule
        SpeechOp (No Transcript) & 2.00 & 4.83 & .908 & 8.1  \\
        \hspace{0.5cm} +ITC & 2.05 (\textcolor{blue}{+0.05}) & 4.85 (\textcolor{red}{+0.02}) & .928 (\textcolor{blue}{+.020}) & 2.9 (\textbf{\textcolor{blue}{-5.2}})  \\
        \hspace{0.5cm} +Speaker Personalization & 2.12 (\textcolor{blue}{+0.07}) & 4.69 (\textbf{\textcolor{blue}{-0.16}}) & .926 (\textcolor{red}{-.002}) & 2.4 (\textbf{\textcolor{blue}{-0.5}}) \\
        \midrule
        SpeechOp (Gold Transcript) & 2.06 & 4.83 & .931 & 2.1  \\
        \bottomrule
        \end{tabular}%
        }
    \end{minipage}%
    \hfill
    \begin{minipage}[t]{0.33\linewidth}
        \centering
        % \captionof{table}{Speech Enhancement Results (Subj.). Avg. MOS \& std. error.} % REMOVED
        % \label{tab:enhancement_results_subj} % REMOVED
        \resizebox{\linewidth}{!}{
        \begin{tabular}{lc}
        % ... subjective table content ...
        \toprule
        Model & MOS $\uparrow$  \\
        \midrule
        Noisy Source Audio & 1.78 {\tiny$\pm$ 0.07} \\
        \midrule
        SGMSE+ & 3.76 {\tiny $\pm$ 0.03} \\
        HiFi-GAN-2 &  3.90 {\tiny $\pm$ 0.04} \\
        \midrule
        SpeechOp (No Transcript) & 3.93 {\tiny $\pm$ 0.04} \\
        SpeechOp-ITC (WhisperX) & 3.89 {\tiny $\pm$ 0.04} \\
        \midrule
        Clean Reference Audio & 4.67 {\tiny $\pm$ 0.02} \\
        \bottomrule
        \end{tabular}
        }
    \end{minipage}
    \vspace{-15pt}
\end{table}

\section{Results and Discussion}
\label{sec:results}

\begin{table*}[h]
\caption{Speaker Separation Evaluation (Subj.). We report the average MOS and the standard error.}
\label{tab:separation_results_mos_dataset_concise_stderr_small}
\scriptsize
\centering
\begin{tabular}{lccccc}
\toprule
Model & LibriMix Clean & LibriMix Noise & WHAMR & WSJ2-Mix & \cellcolor{verylightgray}Total\\
\midrule
Sepformer \cite{chen24h_interspeech} & 3.32 {\tiny$\pm$ 0.07} & 2.95 {\tiny$\pm$ 0.07} & 3.06 {\tiny$\pm$ 0.07} & 3.53 {\tiny$\pm$ 0.07} & \cellcolor{verylightgray}3.22 {\tiny$\pm$ 0.04} \\
DM Sepformer \cite{chen24h_interspeech} & 3.59 {\tiny$\pm$ 0.07} & 2.67 {\tiny$\pm$ 0.07} & 2.53 {\tiny$\pm$ 0.07} & 3.58 {\tiny$\pm$ 0.07} & \cellcolor{verylightgray}3.10 {\tiny$\pm$ 0.04}\\
AC-SIM Sepformer \cite{chen24h_interspeech}& 3.74 {\tiny$\pm$ 0.07} & 2.81 {\tiny$\pm$ 0.07} & 2.53 {\tiny$\pm$ 0.07} & 3.65 {\tiny$\pm$ 0.07} & \cellcolor{verylightgray}3.20 {\tiny$\pm$ 0.04}\\
AC-SIM-ML Sepformer \cite{chen24h_interspeech}& 3.74 {\tiny$\pm$ 0.06} & 3.02 {\tiny$\pm$ 0.07} & 2.64 {\tiny$\pm$ 0.07} & 3.66 {\tiny$\pm$ 0.06} & \cellcolor{verylightgray}3.28 {\tiny$\pm$ 0.04}\\
\midrule
SpeechOp (No Transcript) & 3.86 {\tiny$\pm$ 0.07} & 3.68 {\tiny$\pm$ 0.07} & 2.89 {\tiny$\pm$ 0.08} & 3.77 {\tiny$\pm$ 0.07} & \cellcolor{verylightgray}3.57 {\tiny$\pm$ 0.04}\\
SpeechOp (Gold Transcript) & 4.13 {\tiny$\pm$ 0.06} & 4.21 {\tiny$\pm$ 0.06} & 3.37 {\tiny$\pm$ 0.08} & 3.91 {\tiny$\pm$ 0.06} & \cellcolor{verylightgray}3.92 {\tiny$\pm$ 0.03}\\
\midrule
Mixture & 1.38 {\tiny$\pm$ 0.05} & 1.35 {\tiny$\pm$ 0.04} & 1.39 {\tiny$\pm$ 0.04} & 1.33 {\tiny$\pm$ 0.05} & \cellcolor{verylightgray}1.36 {\tiny$\pm$ 0.02} \\
Clean Target & 4.26 {\tiny$\pm$ 0.06} & 4.48 {\tiny$\pm$ 0.05} & 4.29 {\tiny$\pm$ 0.06} & 4.00 {\tiny$\pm$ 0.06} & \cellcolor{verylightgray}4.25 {\tiny$\pm$ 0.03}\\
\bottomrule
\end{tabular}
\end{table*}
\vspace{-10pt}
\begin{table}[!htp]
\caption{Quantitative Speaker Separation Performance on the WSJ0-2Mix Dataset.}
\label{tab:sep_quant}
\scriptsize
\centering
% \resizebox{\columnwidth}{!}{  
\begin{tabular}{lccccc}
\toprule
Method & SI-SDRi $\uparrow$ & MCD $\downarrow$ & SpBS $\uparrow$ & WER $\downarrow$ \\
\midrule
Sepformer \tablefootnote{We compare Sepformer models in \citet{chen24h_interspeech} since they support speaker separation in multiple acoustic environments.}\cite{chen24h_interspeech} & 11.86 & 1.72 & .929 & 4.4\\
AC-SIM-ML Sepformer \cite{chen24h_interspeech} & 11.80 & 1.55 & .931 & 6.8\\
\midrule
SpeechOp (No Transcript) & 0.23 & 4.11 & .899 & 11.1 \\
SpeechOp (Gold Transcript) & 0.53 & 4.20 & .919 & 5.5  \\
\bottomrule
\end{tabular}%}
\vspace{-10pt}
\end{table}

\para{Text-To-Speech.} To examine the impact of multi-task training on text-to-speech, we evaluate our model's zero-shot TTS performance with 3 second speech prompts. Crucially, we initialize SpeechOp from our TTS Baseline, allowing us to directly assess the impact of multi-task training.

Table~\ref{tab:zeroshot_tts_results} demonstrates that SpeechOp not only preserves but \textit{enhances} zero-shot TTS capabilities. After undergoing multi-task training, SpeechOp improves performance across all MOS metrics and objective speaker similarity compared to the TTS Baseline, with minimal loss of intelligibility. Exposure to tasks like enhancement and separation likely enhance SpeechOp's ability to generalize and generate natural speech across diverse acoustic environments.

Against recent TTS systems of comparable scale, SpeechOp exhibits strong performance, matching or exceeding CLaM-TTS and XTTS on most metrics. Impressively, it also surpasses the larger VoiceCraft model in intelligibility and subjective quality. While DiTTO-TTS, a larger model trained on more diverse data, achieves higher overall scores, SpeechOp's results are highly competitive within its class. Future work will explore scaling SpeechOp to leverage similar large-scale datasets.

Beyond zero-shot TTS, SpeechOp demonstrates state-of-the-art capabilities in speech editing. As shown in Table~\ref{tab:speech_editing_results}, SpeechOp significantly outperforms VoiceCraft across all subjective MOS metrics, despite having fewer parameters. These results validate the robustness of our multi-task approach, as SpeechOp maintains exceptional speech editing performance while supporting multiple tasks.

\para{Speech Enhancement.}
Our Implicit Task Composition (ITC) pipeline integrates ASR transcripts from Whisper with our TC-CFG method (Section~\ref{sec:task_composition}) to guide speech content. We find that our ITC pipeline achieves state-of-the-art content preservation in speech enhancement. As shown in Table~\ref{tab:enhancement_results_combined}, ITC yields a Word Error Rate (WER) of 2.9\%, a 46\% relative reduction over the strong HiFi-GAN-2 baseline, significantly reducing the content loss common with generative models. Our ITC pipeline leverages web-scale knowledge from ASR models without requiring transcriptions for training the enhancement component itself.

Our ITC's transcript guidance method is more flexible than transcript-conditioned S2S models. Such models can struggle when ASR errors create contradictions between acoustic and textual inputs, or their performance may be upper-bounded by the input audio quality if they cannot generatively restore highly corrupted content. Furthermore, they typically lack control over the influence of the transcript versus the acoustics at inference time.
In contrast, TC-CFG (Eq.~\eqref{eq:final_score_function_main}) provides this control through a tunable guidance strength ($\gamma$). This allows SpeechOp to trade-off prioritizing acoustic fidelity or emphasizing content restoration guided by the transcript depending on the application.

Even using Whisper transcripts derived from the noisy source audio, ITC improves content intelligibility over the original audio (WER 2.9\% vs. 3.3\%) and enhancement without transcripts (WER 8.1\%). This suggests TC-CFG effectively balances the acoustic information from the noisy source audio with with the imperfect guidance from ASR transcription. While signal-fidelity metrics often penalize generative outputs, SpeechOp's ITC matches HiFi-GAN-2's subjective quality (Table~\ref{tab:enhancement_results_combined} Right) while delivering superior content accuracy.

SpeechOp also enables novel applications like personalized enhancement by composing enhancement with voice cloning. This composition improves speaker fidelity (MCD, PESQ) and modestly reduces WER. To provide an upper bound, ground-truth transcripts lead to a 2.1\% WER. Across all scenarios, SpeechOp's ITC, with our composition approach, effectively integrates textual guidance for controllable, content-aware speech enhancement.

\para{Speaker Separation.}
Our speaker separation evaluation reveals a notable tension: SpeechOp achieves significantly higher Mean Opinion Scores (MOS) across all datasets (Table~\ref{tab:separation_results_mos_dataset_concise_stderr_small}) yet lower objective signal-level fidelity (e.g., SI-SDRi, MCD on WSJ0-2Mix, Table~\ref{tab:sep_quant}) compared to Sepformer baselines, despite human listeners consistently preferring SpeechOp.
This discrepancy arises from differing methodologies. Traditional models like Sepformer optimize for signal reconstruction via feature masking. In contrast, SpeechOp, as a generative model, prioritizes perceptual quality, generating natural-sounding separated speech not strictly derived from input mixtures. This aligns with observations that signal-level metrics often poorly correlate with human perception of speech quality in generative contexts \cite{erdoganwcb2023tokensplit, chen24h_interspeech}.
Importantly, transcript guidance substantially improves SpeechOp's content preservation, reducing WER from 11.1\% to 5.5\% with ground-truth transcripts. This highlights our framework's ability to effectively leverage textual information for improved separation accuracy while maintaining its perceptual strengths.

\begin{wraptable}{r}{0.58\textwidth}
\caption{\textbf{Task Composition.} We compare our proposed composition formulation (TC-CFG) against averaging the score vectors (TC-Avg). Gold transcripts are used in this ablation.}
\label{tab:comp_abl}
\centering
\scriptsize
\begin{tabular}{lcccc}
\toprule
Model  & PESQ $\uparrow$ & MCD $\downarrow$ & SpBS $\uparrow$ & WER $\downarrow$ \\
\midrule
Noisy Source Audio  & 1.12 & 11.22 & .888 & 3.3 \\
\cmidrule(lr){1-5}
SpeechOp (No Transcript) & 2.00 & 4.83 & .908 & 8.1  \\
\cmidrule(lr){1-5}
SpeechOp (TC-Avg)  & 1.88 & 5.24 & .909 & 3.4 \\
SpeechOp (TC-CFG) (Ours) & 2.06 & 4.83 & .931 & 2.1 \\
$\Delta$ (TC-CFG vs TC-Avg) & \textbf{\textcolor{blue}{+.18}} & \textbf{\textcolor{blue}{-0.42}} & \textbf{\textcolor{blue}{+.022}} & \textbf{\textcolor{blue}{-1.3}} \\
\bottomrule
\end{tabular}
\vspace{-10pt}
\end{wraptable}

\para{Task Composition Ablation.}
We empirically validate our TC-CFG approach by composing SpeechOp's enhancement capability with TTS-based textual guidance from the gold transcripts (Table~\ref{tab:comp_abl}).
The "SpeechOp (No Transcript)" row serves as a baseline, representing the performance of our enhancement model without any textual guidance.
When employing the standard score averaging approach ("SpeechOp (TC-Avg)"), we observe a degradation in signal fidelity metrics compared to the "No Transcript" baseline (e.g. MCD increases from 4.83 to 5.24). This aligns with the intuition from our synthetic simulation (Figure~\ref{fig:toy_sim_combined}c), where averaging with the broader TTS prior can negatively impact the focused prior of the enhancement model. While TC-Avg does improve content preservation (WER 3.4\% vs. 8.1\% for "No Transcript"), this comes at the cost of acoustic quality and signal fidelity.

In contrast, our proposed composition approach, TC-CFG, demonstrates superior performance across all metrics. It not only achieves the best content preservation with a WER of 2.1\% (a 38\% reduction over TC-Avg's 3.4\% WER), but it also \textit{maintains or improves} signal fidelity compared to the "No Transcript" baseline (e.g. PESQ 2.06 vs. 2.00).

These results empirically confirm that our TC-CFG formulation effectively isolates text-conditional guidance without degrading acoustic quality. This allows SpeechOp to leverage knowledge from the TTS model for robust content preservation (low WER) while simultaneously maintaining, and even slightly enhancing, the acoustic quality and speaker characteristics established by the enhancement model. This careful decomposition of task-specific guidance is crucial for enabling effective and high-fidelity task composition in generative speech processing.

\section{Conclusion}
\label{sec:conclusion}

In this work, we addressed a fundamental data disparity between text-to-speech synthesis and speech-to-speech tasks by adapting pre-trained TTS models to enable high-quality speech processing despite limited paired data. Through SpeechOp, we showed that multi-task training and principled task composition preserve TTS capabilities while enabling flexible speech-to-speech processing. Our Implicit Task Composition framework demonstrated how to leverage web-scale speech understanding from discriminative models to achieve state-of-the-art content preservation without parallel data. By bridging the gap between data-rich and data-constrained speech tasks, this work opens new possibilities for unified, scalable speech processing systems.

\bibliography{custom}

\begin{thebibliography}{58}
\providecommand{\natexlab}[1]{#1}
\providecommand{\url}[1]{\texttt{#1}}
\expandafter\ifx\csname urlstyle\endcsname\relax
  \providecommand{\doi}[1]{doi: #1}\else
  \providecommand{\doi}{doi: \begingroup \urlstyle{rm}\Url}\fi

\bibitem[ech()]{echothief2024}
Echothief [dataset].
\newblock \url{http://www.echothief.com/echothief/}.
\newblock Accessed: 2024-03-12.

\bibitem[pro()]{prolific}
Prolific.
\newblock \url{https://www.prolific.co/}.

\bibitem[Bain et~al.(2023)Bain, Huh, Han, and Zisserman]{bain2022whisperx}
M.~Bain, J.~Huh, T.~Han, and A.~Zisserman.
\newblock Whisperx: Time-accurate speech transcription of long-form audio.
\newblock \emph{Proc. Interspeech}, 2023.

\bibitem[Casanova et~al.(2024)Casanova, Davis, G{\"o}lge, G{\"o}knar, Gulea, Hart, Aljafari, Meyer, Morais, Olayemi, et~al.]{casanova2024xtts}
E.~Casanova, K.~Davis, E.~G{\"o}lge, G.~G{\"o}knar, I.~Gulea, L.~Hart, A.~Aljafari, J.~Meyer, R.~Morais, S.~Olayemi, et~al.
\newblock Xtts: a massively multilingual zero-shot text-to-speech model.
\newblock \emph{CoRR}, 2024.

\bibitem[Chen et~al.(2021{\natexlab{a}})Chen, Chai, Wang, Du, Zhang, Weng, Su, Povey, Trmal, Zhang, et~al.]{GigaSpeech2021}
G.~Chen, S.~Chai, G.~Wang, J.~Du, W.-Q. Zhang, C.~Weng, D.~Su, D.~Povey, J.~Trmal, J.~Zhang, et~al.
\newblock Gigaspeech: An evolving, multi-domain asr corpus with 10,000 hours of transcribed audio.
\newblock In \emph{Proc. Interspeech}, 2021{\natexlab{a}}.

\bibitem[Chen et~al.(2024{\natexlab{a}})Chen, Su, Berg-Kirkpatrick, Dubnov, and Jin]{chen24h_interspeech}
K.~Chen, J.~Su, T.~Berg-Kirkpatrick, S.~Dubnov, and Z.~Jin.
\newblock Improving generalization of speech separation in real-world scenarios: Strategies in simulation, optimization, and evaluation.
\newblock In \emph{Proc. Interspeech}, 2024{\natexlab{a}}.

\bibitem[Chen et~al.(2021{\natexlab{b}})Chen, Wang, Chen, Wu, Liu, Chen, Li, Kanda, Yoshioka, Xiao, Wu, Zhou, Ren, Qian, Qian, Wu, Zeng, and Wei]{Chen2021WavLM}
S.~Chen, C.~Wang, Z.~Chen, Y.~Wu, S.~Liu, Z.~Chen, J.~Li, N.~Kanda, T.~Yoshioka, X.~Xiao, J.~Wu, L.~Zhou, S.~Ren, Y.~Qian, Y.~Qian, J.~Wu, M.~Zeng, and F.~Wei.
\newblock Wavlm: Large-scale self-supervised pre-training for full stack speech processing.
\newblock \emph{CoRR}, 2021{\natexlab{b}}.

\bibitem[Chen et~al.(2022)Chen, Hu, and Owens]{chen2022structure}
Z.~Chen, X.~Hu, and A.~Owens.
\newblock Structure from silence: Learning scene structure from ambient sound.
\newblock In \emph{Proc. Conference on Robot Learning}, 2022.

\bibitem[Chen et~al.(2024{\natexlab{b}})Chen, Geng, and Owens]{chen2024images}
Z.~Chen, D.~Geng, and A.~Owens.
\newblock Images that sound: Composing images and sounds on a single canvas.
\newblock \emph{CoRR}, 2024{\natexlab{b}}.

\bibitem[Dubey et~al.(2024)Dubey, Aazami, Gopal, Naderi, Braun, Cutler, Ju, Zohourian, Tang, Golestaneh, and Aichner]{dns_challenge}
H.~Dubey, A.~Aazami, V.~Gopal, B.~Naderi, S.~Braun, R.~Cutler, A.~Ju, M.~Zohourian, M.~Tang, M.~Golestaneh, and R.~Aichner.
\newblock Icassp 2023 deep noise suppression challenge.
\newblock \emph{IEEE Open Journal of Signal Processing}, 2024.

\bibitem[Erdogan et~al.(2023)Erdogan, Wisdom, Chang, Borsos, Tagliasacchi, Zeghidour, and Hershey]{erdoganwcb2023tokensplit}
H.~Erdogan, S.~Wisdom, X.~Chang, Z.~Borsos, M.~Tagliasacchi, N.~Zeghidour, and J.~R. Hershey.
\newblock Tokensplit: Using discrete speech representations for direct, refined, and transcript-conditioned speech separation and recognition.
\newblock In \emph{Proc. Interspeech}, 2023.

\bibitem[Ho and Salimans(2022)]{ho2022classifier}
J.~Ho and T.~Salimans.
\newblock Classifier-free diffusion guidance.
\newblock \emph{CoRR}, 2022.

\bibitem[Ho et~al.(2020)Ho, Jain, and Abbeel]{ho2020denoising}
J.~Ho, A.~Jain, and P.~Abbeel.
\newblock Denoising diffusion probabilistic models.
\newblock \emph{Proc. NeurIPS}, 2020.

\bibitem[Hoogeboom et~al.(2023)Hoogeboom, Heek, and Salimans]{hoogeboom2023simple}
E.~Hoogeboom, J.~Heek, and T.~Salimans.
\newblock simple diffusion: End-to-end diffusion for high resolution images.
\newblock In \emph{Proc. ICML}, 2023.

\bibitem[Hoogeboom et~al.(2024)Hoogeboom, Mensink, Heek, Lamerigts, Gao, and Salimans]{hoogeboom2024simpler}
E.~Hoogeboom, T.~Mensink, J.~Heek, K.~Lamerigts, R.~Gao, and T.~Salimans.
\newblock Simpler diffusion (sid2): 1.5 fid on imagenet512 with pixel-space diffusion.
\newblock \emph{CoRR}, 2024.

\bibitem[Hsu et~al.(2021)Hsu, Bolte, Tsai, Lakhotia, Salakhutdinov, and Mohamed]{hsu2021hubert}
W.-N. Hsu, B.~Bolte, Y.-H.~H. Tsai, K.~Lakhotia, R.~Salakhutdinov, and A.~Mohamed.
\newblock Hubert: Self-supervised speech representation learning by masked prediction of hidden units.
\newblock \emph{IEEE Trans. Audio, Speech, Lang. Process.}, 2021.

\bibitem[Ju et~al.(2024)Ju, Wang, Shen, Tan, Xin, Yang, Liu, Leng, Song, Tang, et~al.]{junaturalspeech}
Z.~Ju, Y.~Wang, K.~Shen, X.~Tan, D.~Xin, D.~Yang, E.~Liu, Y.~Leng, K.~Song, S.~Tang, et~al.
\newblock Naturalspeech 3: Zero-shot speech synthesis with factorized codec and diffusion models.
\newblock In \emph{Forty-first International Conference on Machine Learning}, 2024.

\bibitem[Karras et~al.(2022)Karras, Aittala, Aila, and Laine]{karras2022elucidating}
T.~Karras, M.~Aittala, T.~Aila, and S.~Laine.
\newblock Elucidating the design space of diffusion-based generative models.
\newblock \emph{Advances in neural information processing systems}, 35:\penalty0 26565--26577, 2022.

\bibitem[Kim et~al.(2024)Kim, Lee, Chung, and Cho]{kimclam}
J.~Kim, K.~Lee, S.~Chung, and J.~Cho.
\newblock Clam-tts: Improving neural codec language model for zero-shot text-to-speech.
\newblock In \emph{Proc. ICLR}, 2024.

\bibitem[Kingma and Gao(2023)]{kingma2023understanding}
D.~P. Kingma and R.~Gao.
\newblock Understanding diffusion objectives as the {ELBO} with simple data augmentation.
\newblock In \emph{Proc. NeurIPS}, 2023.

\bibitem[Ko et~al.(2017)Ko, Peddinti, Povey, Seltzer, and Khudanpur]{openslr28}
T.~Ko, V.~Peddinti, D.~Povey, M.~L. Seltzer, and S.~Khudanpur.
\newblock A study on data augmentation of reverberant speech for robust speech recognition.
\newblock In \emph{Proc. ICASSP}, 2017.

\bibitem[Koizumi et~al.(2023{\natexlab{a}})Koizumi, Zen, Karita, Ding, Yatabe, Morioka, Bacchiani, Zhang, Han, and Bapna]{koizumi23_interspeech}
Y.~Koizumi, H.~Zen, S.~Karita, Y.~Ding, K.~Yatabe, N.~Morioka, M.~Bacchiani, Y.~Zhang, W.~Han, and A.~Bapna.
\newblock Libritts-r: A restored multi-speaker text-to-speech corpus.
\newblock In \emph{Proc. Interspeech}, 2023{\natexlab{a}}.

\bibitem[Koizumi et~al.(2023{\natexlab{b}})Koizumi, Zen, Karita, Ding, Yatabe, Morioka, Zhang, Han, Bapna, and Bacchiani]{koizumi2023miipher}
Y.~Koizumi, H.~Zen, S.~Karita, Y.~Ding, K.~Yatabe, N.~Morioka, Y.~Zhang, W.~Han, A.~Bapna, and M.~Bacchiani.
\newblock Miipher: A robust speech restoration model integrating self-supervised speech and text representations.
\newblock In \emph{Proc. WASPAA}, 2023{\natexlab{b}}.

\bibitem[Koizumi et~al.(2023{\natexlab{c}})Koizumi, Zen, Karita, Ding, Yatabe, Morioka, Zhang, Han, Bapna, and Bacchiani]{miipher}
Y.~Koizumi, H.~Zen, S.~Karita, Y.~Ding, K.~Yatabe, N.~Morioka, Y.~Zhang, W.~Han, A.~Bapna, and M.~Bacchiani.
\newblock Miipher: A robust speech restoration model integrating self-supervised speech and text representations.
\newblock In \emph{Proc. WASPAA}, 2023{\natexlab{c}}.

\bibitem[Kumar et~al.(2023)Kumar, Seetharaman, Luebs, Kumar, and Kumar]{kumar2024high}
R.~Kumar, P.~Seetharaman, A.~Luebs, I.~Kumar, and K.~Kumar.
\newblock High-fidelity audio compression with improved rvqgan.
\newblock \emph{Proc. NeurIPS}, 2023.

\bibitem[Kynk{\"a}{\"a}nniemi et~al.(2024)Kynk{\"a}{\"a}nniemi, Aittala, Karras, Laine, Aila, and Lehtinen]{nniemi2024applying}
T.~Kynk{\"a}{\"a}nniemi, M.~Aittala, T.~Karras, S.~Laine, T.~Aila, and J.~Lehtinen.
\newblock Applying guidance in a limited interval improves sample and distribution quality in diffusion models.
\newblock In \emph{The Thirty-eighth Annual Conference on Neural Information Processing Systems}, 2024.
\newblock URL \url{https://openreview.net/forum?id=nAIhvNy15T}.

\bibitem[Le et~al.(2024)Le, Vyas, Shi, Karrer, Sari, Moritz, Williamson, Manohar, Adi, Mahadeokar, et~al.]{le2024voicebox}
M.~Le, A.~Vyas, B.~Shi, B.~Karrer, L.~Sari, R.~Moritz, M.~Williamson, V.~Manohar, Y.~Adi, J.~Mahadeokar, et~al.
\newblock Voicebox: Text-guided multilingual universal speech generation at scale.
\newblock 2024.

\bibitem[Lee et~al.(2024)Lee, Kim, Kim, and Cho]{lee2024ditto}
K.~Lee, D.~W. Kim, J.~Kim, and J.~Cho.
\newblock Ditto-tts: Efficient and scalable zero-shot text-to-speech with diffusion transformer.
\newblock \emph{CoRR}, 2024.

\bibitem[Lemercier et~al.(2023)Lemercier, Richter, Welker, and Gerkmann]{lemercier2023storm}
J.-M. Lemercier, J.~Richter, S.~Welker, and T.~Gerkmann.
\newblock Storm: A diffusion-based stochastic regeneration model for speech enhancement and dereverberation.
\newblock \emph{IEEE Trans. Audio, Speech, Lang. Process.}, 2023.

\bibitem[Liu et~al.(2024)Liu, Le, Vyas, Shi, Tjandra, and Hsu]{liu2024generative}
A.~H. Liu, M.~Le, A.~Vyas, B.~Shi, A.~Tjandra, and W.-N. Hsu.
\newblock Generative pre-training for speech with flow matching.
\newblock In \emph{The Twelfth International Conference on Learning Representations}, 2024.
\newblock URL \url{https://openreview.net/forum?id=KpoQSgxbKH}.

\bibitem[Liu et~al.(2022)Liu, Li, Du, Torralba, and Tenenbaum]{liu2022compositional}
N.~Liu, S.~Li, Y.~Du, A.~Torralba, and J.~B. Tenenbaum.
\newblock Compositional visual generation with composable diffusion models.
\newblock In \emph{Proc. ECCV}, 2022.

\bibitem[Lovelace et~al.(2024{\natexlab{a}})Lovelace, Ray, Kim, Weinberger, and Wu]{lovelace2024simpletts}
J.~Lovelace, S.~Ray, K.~Kim, K.~Q. Weinberger, and F.~Wu.
\newblock Simple-{TTS}: End-to-end text-to-speech synthesis with latent diffusion, 2024{\natexlab{a}}.

\bibitem[Lovelace et~al.(2024{\natexlab{b}})Lovelace, Ray, Kim, Weinberger, and Wu]{lovelace24_interspeech}
J.~Lovelace, S.~Ray, K.~Kim, K.~Q. Weinberger, and F.~Wu.
\newblock Sample-efficient diffusion for text-to-speech synthesis.
\newblock In \emph{Proc. Interspeech}, 2024{\natexlab{b}}.

\bibitem[Lu et~al.()Lu, Zhou, Bao, Chen, Li, and Zhu]{ludpm}
C.~Lu, Y.~Zhou, F.~Bao, J.~Chen, C.~Li, and J.~Zhu.
\newblock Dpm-solver++: Fast solver for guided sampling of diffusion probabilistic models.

\bibitem[Maciejewski et~al.(2020)Maciejewski, Wichern, McQuinn, and Le~Roux]{maciejewski2020whamr}
M.~Maciejewski, G.~Wichern, E.~McQuinn, and J.~Le~Roux.
\newblock Whamr!: Noisy and reverberant single-channel speech separation.
\newblock In \emph{Proc. ICASSP}, 2020.

\bibitem[Panayotov et~al.(2015)Panayotov, Chen, Povey, and Khudanpur]{librispeech}
V.~Panayotov, G.~Chen, D.~Povey, and S.~Khudanpur.
\newblock Librispeech: An asr corpus based on public domain audio books.
\newblock In \emph{Proc. ICASSP}, 2015.

\bibitem[Peebles and Xie(2022)]{peebles2022scalable}
W.~Peebles and S.~Xie.
\newblock Scalable diffusion models with transformers.
\newblock \emph{CoRR}, 2022.

\bibitem[Peebles and Xie(2023)]{peebles2023scalable}
W.~Peebles and S.~Xie.
\newblock Scalable diffusion models with transformers.
\newblock In \emph{Proc. CVPR}, 2023.

\bibitem[Peng et~al.(2024)Peng, Huang, Li, Mohamed, and Harwath]{peng2024voicecraft}
P.~Peng, P.-Y. Huang, S.-W. Li, A.~Mohamed, and D.~Harwath.
\newblock Voicecraft: Zero-shot speech editing and text-to-speech in the wild.
\newblock \emph{CoRR}, 2024.

\bibitem[Pratap et~al.(2020)Pratap, Xu, Sriram, Synnaeve, and Collobert]{pratap2020mls}
V.~Pratap, Q.~Xu, A.~Sriram, G.~Synnaeve, and R.~Collobert.
\newblock Mls: A large-scale multilingual dataset for speech research.
\newblock 2020.

\bibitem[Radford et~al.(2023)Radford, Kim, Xu, Brockman, McLeavey, and Sutskever]{radford2023robust}
A.~Radford, J.~W. Kim, T.~Xu, G.~Brockman, C.~McLeavey, and I.~Sutskever.
\newblock Robust speech recognition via large-scale weak supervision.
\newblock In \emph{Proc. ICML}, 2023.

\bibitem[Richter et~al.(2023)Richter, Welker, Lemercier, Lay, and Gerkmann]{richter2023speech}
J.~Richter, S.~Welker, J.-M. Lemercier, B.~Lay, and T.~Gerkmann.
\newblock Speech enhancement and dereverberation with diffusion-based generative models.
\newblock \emph{IEEE Trans. Audio, Speech, Lang. Process.}, 2023.

\bibitem[Rombach et~al.(2021)Rombach, Blattmann, Lorenz, Esser, and Ommer]{rombach2021highresolution}
R.~Rombach, A.~Blattmann, D.~Lorenz, P.~Esser, and B.~Ommer.
\newblock High-resolution image synthesis with latent diffusion models.
\newblock \emph{CoRR}, 2021.

\bibitem[Roux et~al.(2019)Roux, Wisdom, Erdogan, and Hershey]{rouxwe2019SISDR}
J.~L. Roux, S.~Wisdom, H.~Erdogan, and J.~R. Hershey.
\newblock {SDR} - half-baked or well done?
\newblock In \emph{Proc. ICASSP}, 2019.

\bibitem[Saeki et~al.(2024)Saeki, Maiti, Takamichi, Watanabe, and Saruwatari]{saeki2024speechbertscore}
T.~Saeki, S.~Maiti, S.~Takamichi, S.~Watanabe, and H.~Saruwatari.
\newblock Speechbertscore: Reference-aware automatic evaluation of speech generation leveraging nlp evaluation metrics.
\newblock \emph{CoRR}, 2024.

\bibitem[Salimans and Ho(2022)]{salimans2022progressive}
T.~Salimans and J.~Ho.
\newblock Progressive distillation for fast sampling of diffusion models.
\newblock In \emph{Proc. ICLR}, 2022.

\bibitem[Shen et~al.(2022)Shen, Ju, Tan, Liu, Leng, He, Qin, Bian, et~al.]{shennaturalspeech}
K.~Shen, Z.~Ju, X.~Tan, E.~Liu, Y.~Leng, L.~He, T.~Qin, J.~Bian, et~al.
\newblock Naturalspeech 2: Latent diffusion models are natural and zero-shot speech and singing synthesizers.
\newblock In \emph{Proc. ICLR}, 2022.

\bibitem[Song and Ermon(2019)]{song2019generative}
Y.~Song and S.~Ermon.
\newblock Generative modeling by estimating gradients of the data distribution.
\newblock \emph{Proc. NeurIPS}, 2019.

\bibitem[Su et~al.(2021{\natexlab{a}})Su, Jin, and Finkelstein]{su2021hifigan2}
J.~Su, Z.~Jin, and A.~Finkelstein.
\newblock Hifi-gan-2: Studio-quality speech enhancement via generative adversarial networks conditioned on acoustic features.
\newblock In \emph{Proc. WASPAA}, 2021{\natexlab{a}}.

\bibitem[Su et~al.(2021{\natexlab{b}})Su, Wang, Finkelstein, and Jin]{subandwidth2021}
J.~Su, Y.~Wang, A.~Finkelstein, and Z.~Jin.
\newblock Bandwidth extension is all you need.
\newblock In \emph{Proc. ICASSP}, 2021{\natexlab{b}}.

\bibitem[Subakan et~al.(2021)Subakan, Ravanelli, Cornell, Bronzi, and Zhong]{subakan2021attention}
C.~Subakan, M.~Ravanelli, S.~Cornell, M.~Bronzi, and J.~Zhong.
\newblock Attention is all you need in speech separation.
\newblock In \emph{Proc. ICASSP}, 2021.

\bibitem[Traer and McDermott(2016)]{ir_survey}
J.~Traer and J.~H. McDermott.
\newblock Statistics of natural reverberation enable perceptual separation of sound and space.
\newblock \emph{Proceedings of the National Academy of Sciences}, 2016.

\bibitem[Valle et~al.(2025)Valle, Badlani, Kong, Lee, Goel, Kim, Santos, Dai, Gururani, AlJa'fari, Liu, Shih, Ping, and Catanzaro]{valle2025fugatto}
R.~Valle, R.~Badlani, Z.~Kong, S.-g. Lee, A.~Goel, S.~Kim, J.~F. Santos, S.~Dai, S.~Gururani, A.~AlJa'fari, A.~Liu, K.~Shih, W.~Ping, and B.~Catanzaro.
\newblock Fugatto 1: Foundational generative audio transformer opus 1.
\newblock In \emph{Proc. ICLR}, 2025.

\bibitem[Wang et~al.(2024)Wang, Thakker, Chen, Kanda, Eskimez, Chen, Tang, Liu, Li, and Yoshioka]{wang2024speechx}
X.~Wang, M.~Thakker, Z.~Chen, N.~Kanda, S.~E. Eskimez, S.~Chen, M.~Tang, S.~Liu, J.~Li, and T.~Yoshioka.
\newblock Speechx: Neural codec language model as a versatile speech transformer.
\newblock \emph{IEEE Trans. Audio, Speech, Lang. Process.}, 2024.

\bibitem[Xue et~al.(2022)Xue, Barua, Constant, Al-Rfou, Narang, Kale, Roberts, and Raffel]{xue2022byt5}
L.~Xue, A.~Barua, N.~Constant, R.~Al-Rfou, S.~Narang, M.~Kale, A.~Roberts, and C.~Raffel.
\newblock Byt5: Towards a token-free future with pre-trained byte-to-byte models.
\newblock \emph{Transactions of the Association for Computational Linguistics}, 2022.

\bibitem[Yang et~al.(2023)Yang, Tian, Tan, Huang, Liu, Chang, Shi, Zhao, Bian, Wu, et~al.]{yang2023uniaudio}
D.~Yang, J.~Tian, X.~Tan, R.~Huang, S.~Liu, X.~Chang, J.~Shi, S.~Zhao, J.~Bian, X.~Wu, et~al.
\newblock Uniaudio: An audio foundation model toward universal audio generation.
\newblock \emph{arXiv preprint arXiv:2310.00704}, 2023.

\bibitem[Yang et~al.(2024)Yang, Su, Kim, and Jin]{yang24h_interspeech}
H.~Yang, J.~Su, M.~Kim, and Z.~Jin.
\newblock Genhancer: High-fidelity speech enhancement via generative modeling on discrete codec tokens.
\newblock In \emph{Proc. Interspeech}, 2024.

\bibitem[Zen et~al.(2019)Zen, Dang, Clark, Zhang, Weiss, Jia, Chen, and Wu]{zen2019libritts}
H.~Zen, V.~Dang, R.~Clark, Y.~Zhang, R.~J. Weiss, Y.~Jia, Z.~Chen, and Y.~Wu.
\newblock Libritts: A corpus derived from librispeech for text-to-speech.
\newblock 2019.

\end{thebibliography}
\bibliographystyle{abbrvnat}

% % \input{sections/checklist}
\appendix
\onecolumn

\section{Subjective Study Details}
\para{Methodology.} Our studies used native English speakers on Prolific to measure naturalness, quality, voice similarity, and style similarity on a 1-5 scale, for text-to-speech synthesis based on a reference speech sample. For our speech processing tasks, we measure the quality of the audio sample. We also included a flag for unintelligible content, though no samples were ultimately flagged by a majority of raters.

\para{Quality Control.} To filter out unreliable ratings for the TTS and speech editing studies, we used two types of hidden validation tests. The first was a mismatched speaker test (different but real speakers for reference and sample); if a participant rated speaker similarity > 3, their ratings were discarded. The second was an identical pair test; if any attribute was rated < 4, their ratings were discarded. For the speech processing tasks, we conducted similar validation tests with the clean and noisy audio samples.

\para{Participant details.} For all of our subjective tests, each worker rated ~30 samples, including 4 validation tests. Our TTS study involved 288 unique workers rating 80 utterances per method. Our speech editing study involved 151 unique workers rating 100 utterances per method. Our enhancement study involved 236 unique workers rating 96 utterances per method.

\para{Compensation.} For our listening experiments, participants were compensated at a rate of \$15/hour, which is above the platform's recommendation.

\section{Simulation Study Implementation Details}

For our guidance comparison simulation study on the 1D Gaussian Mixture Model, we provide detailed implementation specifics to ensure reproducibility.

\subsection{Tractable Ground Truth Computation}

The 1D GMM setting enables exact computation of all relevant quantities, providing a controlled environment for comparing guidance strategies. Both the conditional score function, $\nabla_{x_t} \log p_t(x_t | y)$, and guidance term, $\nabla_{x_t} \log p_{\text{TTS}}(y|x_t)$, can be computed analytically. 

\subsection{Experimental Parameters}

Our synthetic experiments use the configuration detailed in Table \ref{tab:guidance_simulation_params}.

\begin{table}[h]
\centering
\caption{Guidance Comparison Simulation Parameters}
\label{tab:guidance_simulation_params}
\begin{tabular}{ll}
\toprule
\textbf{Parameter} & \textbf{Value} \\
\midrule
\multicolumn{2}{l}{\textit{\textbf{Base Parameters}}} \\
\midrule
\multicolumn{2}{l}{\textit{TTS Distribution (Generic Speech)}} \\
Component means & $\mu_0 = -4.0$, $\mu_1 = 4.0$ \\
Component std. devs. & $\sigma_0 = \sigma_1 = 0.9$ \\
Component weights & $w_0 = w_1 = 0.5$ \\
Target component & $y = 0$ \\
\midrule
\multicolumn{2}{l}{\textit{Enhancement Transforms}} \\
Mean shift & $\Delta\mu = 2.0$ \\
Variance reduction factor & $\gamma = 4$ \\
Imperfect model bias & $\epsilon = 0.4$ \\
Imperfect model variance inflation & $\beta = 1.8$ \\
\midrule
\multicolumn{2}{l}{\textit{\textbf{Derived Parameters}}} \\
\midrule
\multicolumn{2}{l}{\textit{True Enhanced Speech}} \\
Mean & $\mu_0 + \Delta\mu = -2.0$ \\
Std. dev. & $\sigma_0/\gamma = 0.23$ \\
\midrule
\multicolumn{2}{l}{\textit{Imperfect Enhancement Model}} \\
Mean & $\mu_0 + \Delta\mu + \epsilon = -1.6$ \\
Std. dev. & $\beta \cdot \sigma_0/\gamma = 0.41$ \\
\midrule
\multicolumn{2}{l}{\textit{Sampling Parameters}} \\
Number of samples & 5000 \\
Number of timesteps & 200 \\
Initial noise level & $\sigma_{\text{max}} = 80$ \\
Final noise level & $\sigma_{\text{min}} = 0.005$ \\
\midrule
\multicolumn{2}{l}{\textit{Guidance Parameters}} \\
CFG guidance strength & $\rho = 10^4$ \\
Score averaging weight & $\alpha = 0.5$ \\
\bottomrule
\end{tabular}
\end{table}

\subsection{Guidance Strategy Comparison}

Our simulation compares three fundamental approaches to combining TTS and speech enhancement models:

\textbf{No Guidance:} Uses only the imperfect enhancement model's score function, representing current single-task approaches:
$$s_{\text{total}} = s_{\text{enh}}(x_t, \sigma_t)$$

\textbf{CFG-Style Guidance:} Augments the enhancement model with discriminative guidance from the TTS model using classifier-free guidance:
$$s_{\text{total}} = s_{\text{enh}}(x_t, \sigma_t) + \rho \cdot \nabla_{x_t} \log p_{\text{TTS}}(y|x_t)$$
where the guidance term $\nabla_{x_t} \log p_{\text{TTS}}(y|x_t)$ leverages the TTS model's ability to distinguish content-matching samples.

\textbf{Score Averaging:} Linearly combines the enhancement model score with the true conditional TTS score:
$$s_{\text{total}} = (1-\alpha) \cdot s_{\text{enh}}(x_t, \sigma_t) + \alpha \cdot s_{\text{TTS}}(x_t, \sigma_t | y)$$
This approach directly mixes the score functions from both models.

\subsection{Noise Schedule and Sampling}

We employ a log-linear interpolation for noise levels:
$\sigma_t = \exp\left(\frac{t}{T} \log(\sigma_{\text{final}}) + \frac{T-t}{T} \log(\sigma_{\text{init}})\right)$

The update step follows the variance exploding diffusion formulation \cite{karras2022elucidating}:
$x_{t+1} = x_t + (\sigma_t^2 - \sigma_{t+1}^2) \cdot s_{\text{total}} + \sqrt{\sigma_t^2 - \sigma_{t+1}^2} \cdot \epsilon$
where $\epsilon \sim \mathcal{N}(0, 1)$.

\subsection{Evaluation Metrics}

We evaluate the final samples using KL divergence computed between the empirical distribution of generated samples and the true enhanced speech distribution, representing the ideal outcome.

\section{Audio AutoEncoder}
\label{app:autoencoder}
For efficient latent diffusion modeling, we develop an autoencder based on DAC \citep{kumar2024high} but with a continuous variational bottleneck instead of residual vector quantization. For 48 kHz input audio $\mathbf{y} \in \mathbb{R}^{1\times T}$, the encoder $E$ maps to latent representations $\mathbf{x}_0 = E(\mathbf{y})$ with dimensions $\mathbb{R}^{C\times L}$, where $C=64$ is the latent channel dimension and $L$ is the temporal dimension downsampled by a factor of 1200 (resulting in a 40 Hz latent representation). The decoder $D$ mirrors this architecture to reconstruct the waveform.

The encoder's output is transformed into latent variables through a variational bottleneck that models the approximate posterior $q(\mathbf{z}|\mathbf{y})$:
\begin{equation}
    \mathbf{z} = \boldsymbol{\mu} + \boldsymbol{\sigma} \odot \boldsymbol{\epsilon}, \quad \text{where} \quad \boldsymbol{\epsilon} \sim \mathcal{N}(0, \mathbf{I})
\end{equation}

The model is trained to minimize reconstruction loss and KL divergence:
\begin{equation}
    \mathcal{L}_{\text{AE}} = \mathbb{E}_{\mathbf{y}}[\|\mathbf{y} - \hat{\mathbf{y}}\|_1] + \lambda_{\text{KL}}\mathcal{L}_{\text{KL}}
\end{equation}
where $\lambda_{\text{KL}} = 0.1$ balances the objectives. We also employ adversarial training with a complex STFT discriminator following DAC to improve reconstruction quality.

\section{Acoustic Simulation}
\label{app:simulation}
First, we randomly select a clean speech sample and apply random equalization and compression. Background noise is then added at a signal-to-noise ratio (SNR) of $-10$ to $30$ dB. We randomly apply reverberation using impulse response (IR) samples. Additional degradation like random bandlimiting down to 1kHz is applied to simulate input at various sample rates. We dynamically generate training pairs during training to increase diversity of the degradation combinations. 

We trained the models with public datasets at 44.1k sample rate. The clean speech data is sourced from LibriTTS-R \citep{koizumi23_interspeech} and upsampled to 44.1k sample rate via bandwidth extension \citep{subandwidth2021}. The noise samples include the  DNS Challenge \citep{dns_challenge} and SFS-Static-Dataset \citep{chen2022structure}. The impulse response (IR) data includes MIT IR Survey \citep{ir_survey}, EchoThief \citep{echothief2024}, and OpenSLR28 \citep{openslr28}.

\section{Architecture and Training Details}
\label{appendix:arch_details}

\subsection{Model Architecture}
We present our model architecture details in Table~\ref{tab:arch_params}. For our transfer learning experiments and architecture ablation, we utilize a smaller version of SpeechOp with 12 DiT layers and 6 encoder layers. We also incorporate dense connections \citep{lee2024ditto}, a position-aware cross-attention mechanism (\citep{lovelace24_interspeech}), and append 8 register tokens to process global information \citep{lovelace24_interspeech}. We condition on the learnable task embedding by summing it with the timestep embedding that given to the DiT network.

\begin{table}[h]
\caption{\textbf{SpeechOp Architecture Parameters}}
\label{tab:arch_params}
\centering
\begin{tabular}{lc}
\toprule
Parameter & Value \\
\midrule
\multicolumn{2}{l}{\textit{Diffusion Transformer }} \\
Audio Latent Dimension & 64 \\
Model Dimension & 1024 \\
Feed-forward Dimension & 3072 \\
Attention Heads & 8 \\
Number of Layers & 20 \\
Dropout & 0.1 \\
\midrule
\multicolumn{2}{l}{\textit{Audio Encoder}} \\
Model Dimension & 768 \\
Feed-forward Dimension & 2304 \\
Number of Layers &  8 \\
\midrule
\multicolumn{2}{l}{\textit{Common Components}} \\
Position Encoding & Rotary \\
Layer Normalization & AdaLN ($\epsilon$=1e-5) \\
Activation & SwiGLU \\
Text Encoder & ByT5-base \\
\bottomrule
\end{tabular}
\end{table}

\subsection{Training Configuration}
All model training is distributed across 32 Nvidia A100s. Training proceeds in two stages:

\paragraph{Stage 1: TTS Pre-training}
Model is trained for 400K iterations with a batch size of 4 per GPU. We use AdamW optimization with learning rate 2e-4 and weight decay 0.1. Training employs 4000 warmup steps and we perform two steps of gradient accumulation. 

\paragraph{Stage 2: Multi-task Fine-tuning}
Starting from the pre-trained TTS model, we extend the encoder to 8 layers and train for an additional 200K iterations. We use a lower learning rate of 1e-4 and weight decay of 0.01, with two steps of gradient accumulation. Batch sizes are 4 for TTS and 8 for speech-to-speech tasks per GPU.

\begin{table}[h]
\caption{\textbf{Multi-task Training Weights and Prompt Probabilities}}
\label{tab:task_weights}
\centering
\begin{tabular}{lcc}
\toprule
Task & Weight & Prompt Probability \\
\midrule
Speech Enhancement & 3.0 & 0.1 \\
Speaker Separation & 3.0 & 0.9 \\
Noise Isolation & 1.0 & 0.1 \\
Acoustic Matching & 1.0 & 0.9 \\
Speech Isolation & 1.0 & 0.1 \\
\bottomrule
\end{tabular}
\end{table}

Both stages use a shifted cosine noise schedule (scale=0.5) \citep{hoogeboom2023simple,lovelace2024simpletts} with sigmoid loss weighting (bias=-2.5) \citep{hoogeboom2024simpler}, mixed precision (bfloat16), and distributed data parallel (DDP) training.

\subsection{Sampling Configuration}

We use the SDE-DPM-Solver++(2M) as described in \citet{ludpm} for sampling. We utilize 256 inference steps with a schedule that is linear in logSNR. For speech-to-speech tasks, we utilize classifier-free guidance \citep{ho2022classifier} with a strength of 1.5. For zero-shot TTS we use guidance scale of 3.0 for the transcript and prompt conditioning information. For speech editing, we use a guidance scale of 2.0 for the transcript and prompt. 

For zero-shot TTS and speech editing, our non-autoregressive approach requires determining the output duration before generation. We estimate this by first computing the speaking rate (phones per second) from the reference speech prompt. For zero-shot TTS, we then multiply this rate by the phoneme count of the target transcript to determine the output duration. For speech editing, we preserve the original duration for unedited regions and apply the same rate-based estimation for edited segments. We found this simple duration modeling approach sufficient for maintaining natural speaking rates aligned with the reference speaker's style.

For task composition, we can control the guidance strength in the same way. Higher guidance values enforce stronger conditioning at the cost of potentially conflicting the the other task. We use a scale of $1.5$ in our composition experiments. We find that TTS guidance is only necessary for resolving details in modest-to-high SNR regimes, so we enable it for logSNR ranges greater than -1.0 \citep{nniemi2024applying}.

\section{Source Audio Conditioning Ablation}
\begin{table}[h]
\caption{\textbf{Source Audio Conditioning Ablation.} We train an ablation model that conditions on the source sequence vectors with a cross-attention mechanism instead of our frame-wise mixing. For this experiment we use a 
}
\label{tab:comp_abl}
\centering
\begin{tabular}{lcccc}
\toprule
Model  & PESQ $\uparrow$ & MCD $\downarrow$ & SpBS $\uparrow$ & WER $\downarrow$ \\
\midrule
Noisy Source Audio  & 1.12 & 11.22 & .888 & 3.3 \\
\midrule
SpeechOp-Small (Cross-attention)  & 1.18 & 15.4 & .751 & $>$100 \\
w/ chunking  & 1.88 & 4.98 & .900 & 9.6 \\
\midrule
SpeechOp-Small (Framewise-Mixing) (Ours) & 1.96 & 4.86 & .902 & 8.8 \\
\bottomrule
\end{tabular}
\end{table}

\label{sec:ablations}
Using 12-layer models, we compare our framewise mixing strategy against a cross-attention based approach for conditioning on source audio. Table~\ref{tab:comp_abl} shows that the cross-attention variant fails catastrophically when processing sequences other than its 5-second training length (WER $>$ 100\%). Even with explicit padding and chunking to account for this, it shows degraded performance across all metrics. In contrast, our framewise mixing approach generalizes naturally to arbitrary sequence lengths while achieving better quality (PESQ 1.96 vs 1.88), lower distortion (MCD 4.86 vs 4.98), and improved content preservation (WER 8.8\% vs 9.6\%). These results suggest that framewise mixing provides a more robust foundation for speech-to-speech processing, likely due to the explicit frame-level correspondence between source and target audio.

\section{Task Composition Derivation}
\label{app:task_composition_derivation}

Here we present the detailed derivation of our task composition approach. Our goal is to estimate the conditional score function $\nabla_{\mathbf{z}_t} \log p(\mathbf{z}_t|y,w)$, where $\mathbf{z}_t$ is the noisy latent, $y$ is the noisy source audio, and $w$ is the text transcript.

Starting with Bayes' rule, we can decompose the joint conditional probability:
\begin{align}
    \nabla_{\mathbf{z}_t} \log p(\mathbf{z}_t|y,w) &= \nabla_{\mathbf{z}_t} \log \frac{p(y,w|\mathbf{z}_t)p(\mathbf{z}_t)}{p(y,w)} \nonumber \\
    &=  \nabla_{\mathbf{z}_t} \log p(y,w|\mathbf{z}_t) + \nabla_{\mathbf{z}_t} \log p(\mathbf{z}_t) - \nabla_{\mathbf{z}_t} \log p(y,w) \nonumber \\
    &=  \nabla_{\mathbf{z}_t} \log p(y,w|\mathbf{z}_t) + \nabla_{\mathbf{z}_t} \log p(\mathbf{z}_t),
    \label{eq:bayes_rule_appendix}
\end{align}
where we drop the term $\nabla_{\mathbf{z}_t} \log p(y,w)$ as it is independent of $\mathbf{z}_t$.

We introduce a conditional independence assumption: given the noisy latent $\mathbf{z}_t$, the textual transcript $w$ is independent of the noisy source audio $y$. That is:
\begin{equation}
    p(y,w|\mathbf{z}_t) = p(y|\mathbf{z}_t)p(w|\mathbf{z}_t)
    \label{eq:conditional_independence}
\end{equation}

This assumption is reasonable at modest-to-high signal-to-noise ratios where the latent representation effectively captures the salient information from both modalities. Substituting Equation~\eqref{eq:conditional_independence} into Equation~\eqref{eq:bayes_rule_appendix}:
\begin{align}
    \nabla_{\mathbf{z}_t} \log p(\mathbf{z}_t|y,w) &= \nabla_{\mathbf{z}_t} \log p(y|\mathbf{z}_t)p(w|\mathbf{z}_t) + \nabla_{\mathbf{z}_t} \log p(\mathbf{z}_t) \nonumber \\
    &= \nabla_{\mathbf{z}_t} \log p(y|\mathbf{z}_t) + \nabla_{\mathbf{z}_t} \log p(w|\mathbf{z}_t) + \nabla_{\mathbf{z}_t} \log p(\mathbf{z}_t).
    \label{eq:composition_derivation_appendix}
\end{align}

For the term $\nabla_{\mathbf{z}_t} \log p(w|\mathbf{z}_t)$, we can apply Bayes' rule again. Following the classifier-free guidance approach of \citet{ho2022classifier}, this can be expressed in terms of conditional and unconditional TTS score functions:
\begin{equation}
    \nabla_{\mathbf{z}_t} \log p(w|\mathbf{z}_t) = \nabla_{\mathbf{z}_t} \log p(\mathbf{z}_t|w) - \nabla_{\mathbf{z}_t} \log p(\mathbf{z}_t).
    \label{eq:classifier_free_guidance_appendix}
\end{equation}

Substituting Equation~\eqref{eq:classifier_free_guidance_appendix} into Equation~\eqref{eq:composition_derivation_appendix}, and noting that $\nabla_{\mathbf{z}_t} \log p(y|\mathbf{z}_t) + \nabla_{\mathbf{z}_t} \log p(\mathbf{z}_t) = \nabla_{\mathbf{z}_t} \log p(\mathbf{z}_t|y)$, we obtain:
\begin{align}
    \nabla_{\mathbf{z}_t} \log p(\mathbf{z}_t|y,w) &= \nabla_{\mathbf{z}_t} \log p(\mathbf{z}_t|y) + (\nabla_{\mathbf{z}_t} \log p(\mathbf{z}_t|w) - \nabla_{\mathbf{z}_t} \log p(\mathbf{z}_t)) \nonumber \\
    &\approx \mathbf{s}^{\text{enh}}_{\theta}(\mathbf{z}_t|y) + (\mathbf{s}^{\text{tts}}_{\theta}(\mathbf{z}_t|w)- \mathbf{s}^{\text{tts}}_{\theta}(\mathbf{z}_t)),
    \label{eq:final_score_function_appendix}
\end{align}
where $\mathbf{s}^{\text{enh}}_{\theta}(\mathbf{z}_t|y)$ and $\mathbf{s}^{\text{tts}}_{\theta}(\mathbf{z}_t|w)$ represent the score networks for enhancement and TTS tasks, respectively.

This derivation shows how our approach naturally combines the enhancement and TTS score functions while avoiding conflicts between their unconditional priors. The enhancement term guides the denoising process while the TTS term provides content alignment through classifier-free guidance.

\end{document}